\documentclass[prb,reprint,twocolumn,showpacs,superscriptaddress]{revtex4-1}
\usepackage{amsmath,amssymb,bm,mathrsfs,graphicx, braket, times}
\usepackage[colorlinks=true,citecolor=blue,linkcolor=blue]{hyperref}
\usepackage[all,cmtip]{xy}

\renewcommand{\vec}[1]{{\bf{#1}}}

\newcommand{ \f}[2]{f_{#1  #2} }
\newcommand{ \fd}[2]{f_{#1  #2}^{\dagger} }

\begin{document}
\title{Filling-Enforced Quantum Band Insulators in Spin-Orbit Coupled Crystals}

\author{Hoi Chun Po}
\affiliation{University of California, Berkeley, California 94720.}

\author{Haruki Watanabe}
\affiliation{University of California, Berkeley, California 94720.}

\author{Michael P. Zaletel}
\affiliation{Station Q, Microsoft Research, Santa Barbara, California, 93106-6105}

\author{Ashvin Vishwanath}
\affiliation{University of California, Berkeley, California 94720.}
\affiliation{Materials Science Division, Lawrence Berkeley National Laboratories, Berkeley CA 94720}

\begin{abstract}
While band  insulators are usually described in wavevector space in terms of fully filled bands, they are sometimes also described in terms of a  complementary Wannier picture in which electrons occupy localized, atom-like orbitals. Under what conditions does the latter picture break down? The presence of irremovable quantum entanglement between different sites can obstruct a  localized orbital description, which occurs in systems like Chern and topological insulators. We collectively refer to such states as Quantum Band Insulators (QBIs).  Here we report the theoretical discovery of a filling-enforced QBI - that is, a free electron insulator in which the band filling is smaller than the minimum number dictated by the atomic picture. Consequently such insulators have no representation in terms of filling localized orbitals and must be QBIs.  This is shown to occur in models of certain cubic crystals with non-symmorphic space groups. Like topological insulators, filling-enforced QBIs require spin-orbit coupling. However, in contrast, they do not typically exhibit protected surface states. Instead their nontrivial nature is revealed by studying the quantum entanglement of their ground state wavefunction.
\end{abstract}

\maketitle

\section{Introduction}
\label{sec:intro}
An early triumph of quantum mechanics was the realization that the metallic or insulating behavior of crystalline systems could be understood in terms of the partial or full filling of energy bands in wavevector space. For a filled band, a complementary, real space  viewpoint has  electrons localized at points in the unit cell, which in conjunction with the nuclear cores effectively lead to atomic picture  of the insulating state. This was formalized by the work of Wannier et al.~\cite{Resta} which described a procedure to go from fully filled bands to a collection of symmetric and exponentially localized real-space functions, Wannier functions, akin to a collection of isolated atomic orbitals. Whenever such a description is applicable, the system can be understood as an {\em atomic insulator} (AI). 
In order to preserve crystalline symmetries, the centers of these orbitals occupy `Wyckoff positions' in the unit cell and transform into each other under the space group (SG).

However, not all band insulators are AI, meaning they do not admit such a real space depiction - we refer to such insulators as `quantum band insulators.' The best known example is the Chern insulator in two spatial dimensions~\cite{TKNN,Haldane} for which there is an obstruction to constructing localized Wannier functions~\cite{Panati, Brouder}. The obstruction stems  from their nontrivial band topology - a collection of localized atomic orbitals cannot produce emergent chiral edge states. Similarly, topological insulators (TIs) present an obstruction to realizing localized Wannier states that {\em preserve} time reversal (TR) symmetry~\cite{VanderbiltPRB}. Here we report the theoretical discovery of a `filling-enforced' Quantum Band Insulator (feQBI) that shares several features in common with 3D TIs~\cite{TI2013} - including (i) the absence of localized Wannier functions that preserve symmetry, and hence no real space picture of the insulating state and (ii) realization in a free electron Hamiltonian in which spin-orbit coupling is a crucial ingredient. 
In contrast to TIs, however, the feQBI is \emph{required} by the electron filling $\nu$, i.e., the number of electrons per unit cell.
A TI can generally be tuned, preserving symmetries but with a bulk gap closing, into a trivial AI at the same electron filling by modifying the Hamiltonian, while for the feQBI the gap, symmetries, and electron density alone forbid a competing AI state.
Specifically, the feQBI is realized at a filling smaller than twice the minimum number of Wyckoff positions  required to respect all SG symmetries.
SG symmetries play a key role for feQBIs, akin to topological crystalline insulators (TCI) ~\cite{FuPRLTCI}. 
For example, we will show that in SG No.~199 with spin-orbit coupling (SOC), a feQBI occurs at a filling of 4 electrons per unit cell. However, based on the atomic picture the minimum AI filling is 8. 
In fact, feQBIs can only occur in 4  out of the 230 space groups (No.~199, 214, 220 and 230)  all of which are non-symmorphic, cubic crystals.
Since cubic symmetry is always broken at a surface, feQBIs do not necessarily have protected surface states - their nontrivial character will be revealed instead via a suitably engineered entanglement-based approach.

At first sight, the feQBI may seem related to a Mott insulator, which also occurs at `fractional'  filling. However, we emphasize the feQBI is a state of noninteracting electrons, unlike Mott insulators. Furthermore, general interacting arguments which determine allowed  fillings for  band insulators, like that of Hastings-Oshikawa-Lieb-Schultz-Mattis~\cite{Lieb1961, Oshikawa2000, Hastings2004} and their generalizations to non-symmorphic crystals~\cite{Sid2013, Roy} with spin-orbit coupling~\cite{us},  correctly predict 4 as the minimum allowed filling for a band insulator with space group No.~199.

\section{Atomic versus band insulators}

In this work we focus on systems of \emph{non-interacting} spin-1/2 electrons with particle number conservation, TR and SG symmetries.
SOC plays an important role, so there is no spin-rotational invariance.
For brevity, we set all lattice constants to $1$.

Given an SG $\mathcal{G}$, symmetry-related points in space form a crystallographic orbit, and the collection of connected orbits is called a `Wyckoff position'~\cite{ITC}.
Each Wyckoff position $\mathcal W_{w}^{\mathcal G}$ is labeled by a `Wyckoff letter' ($w=$a,b,\dots), ordered by decreasing degree of site symmetry.
A crystallographic orbit in $\mathcal W_{w}^{\mathcal G}$ has `$|\mathcal W_{w}^{\mathcal G}|$' points within each unit cell.
Wyckoff positions with higher site-symmetry have smaller $|\mathcal W_{w}^{\mathcal G}|$. 

SGs are  classified as being either symmorphic or non-symmorphic (NS).
An SG is symmorphic if one can pick an origin `o' for which all symmetry elements can be decomposed into a point group operation which leaves `o' invariant followed by a lattice translation. 
By construction, the crystallographic orbit of `o' is the Bravais lattice, so $|\mathcal W_{\text{a}}^{\text{S}}|=1$.
By definition such an origin is absent in a NS SG, and hence $|\mathcal W_{\text{a}}^{\mathcal G}|\geq 2$ if $\mathcal G$ is NS.

Wyckoff positions are intimately related to AIs, which are real space specification of a system. In the strict AI limit, electrons are localized to a set of points, though the points need not coincide with the real atomic sites in the problem.
Nonetheless, to respect SG symmetries the set of points must correspond to a collection of crystallographic orbits, each in some Wyckoff position, i.e.  the set of points constitute the lattice points of a SG-symmetric lattice.
Since TR has to be respected locally, the system can be an AI only when the number of electrons per point is even.
Note that in a `valence bond' state, the electron positions can be smoothly deformed to the center of mass of each valence bond, which reduces to the above point-like picture.
AIs are therefore only possible at the `atomic fillings' $\nu^{ \text{AI}}=2 \sum_{\{ w\}} |\mathcal W_{w}^{\mathcal G} |$ for some set of Wyckoff positions $\{ w\}$. 
As $ \mathcal W^{\mathcal G}_{\text{a}}$ always has maximal site symmetry, it is always (one of) the position(s) with smallest $|\mathcal W^{\mathcal G}_{w}|$. 
Hence the minimum filling for an AI is  $\nu^{ \text{AI}}_{\text{min}}=2 |\mathcal W_{\text{a}}^{\mathcal G} |$. For symmorphic crystals, $\nu^{ \text{AI}}_{\text{min}} = 2$, while for NS crystals  $\nu^{ \text{AI}}_{\text{min}}\geq 4$.

Free electrons in a periodic potential are also conventionally studied in  momentum space via electronic band structures.
SGs constrain the properties of the bands since they are required to furnish little group irreducible representations (irreps) at high symmetry momenta.
It is well known that bands can be forced to become degenerate due to the dimensionality of the irreps, or due to time-reversal pairing at TR invariant momenta~\cite{Zak2001}.
In addition, the irreps at two end points of a high-symmetry line are not necessarily compatible, and such `compatibility relations' can give further constraints on the possibility of isolating a set of bands. 

The minimal filling for forming a band insulator,  $\nu^{\text{Band}}_{\text{min}}$, is identical to the smallest number of connected bands that can be isolated from all others.
In particular, the class of free electron problems contains those with no hopping between sites.
The ground state of such a Hamiltonian is a strict AI, and hence $\nu^{\text{Band}}_{\text{min}}\leq \nu^{\text{AI}}_{\text{min}}$. In fact, for all SGs one can show that equality holds for spinless, TR-symmetric electrons~\cite{us3}. 

Relaxing the condition on spin-rotation invariance, can the minimal filling for a band insulator be strictly smaller than that for an AI, i.e. can $\nu_{\text{min}}^{\text{Band}} < \nu_{\text{min}}^{\text{AI}}$? 
Here we answer this question in the affirmative: there exist SGs for which TR-symmetric, SOC band insulators can appear at a filling smaller than any AI. A band insulator realized at such sub-atomic filling is \emph{necessarily} non-atomic, and hence the name feQBI.

\section{Example of a feQBI}
\label{sec:feQBI_Ex}
The feQBI examples were found by a systematic study of all the 230 SGs. A particularly interesting observation is that for \emph{almost} all SGs, the number of sites required to furnish any Wyckoff positions is always an integer multiple of $|\mathcal W^{\mathcal G}_{\text{a}}|$, which implies AIs are only possible at fillings $\nu^{\text{AI}}= n \, \nu^{\text{AI}}_{\text{min}}$ for some integer $n$. There are, however, \emph{four} exceptions: No.~199, 214, 220 and 230 (Table \ref{tab:WyckoffPos}). We term such SGs as `Wyckoff-mismatched'.

The simplest of the four Wyckoff-mismatched SG is No.~199, which has $|\mathcal W^{\text{199}}_{\text{a}}|=4$ and therefore $\nu^{\text{AI}}_{\text{min}}=8$. Yet, by studying the spinful little group irreps (Appendix \ref{sec:a199}) and compatibility relations, we discovered a band insulator can be formed at a filling $\nu = 4$, i.e. $\nu^{\text{Band}}_{\text{min}}=4<\nu^{\text{AI}}_{\text{min}}$ and hence proving the existence of feQBI.
As a concrete example, we provide here a simple tight-binding model for this.
We consider a set of sites in $\mathcal W^{\text{199}}_{\text{b}}$, which contains six sites in each primitive unit cell. On the $l$-th ($l=1,2,\dots,6$) site of the cell at $\vec x$, we consider an s-orbital with electron operator $f^{l s \dagger}_{\vec x}$, where $s =\uparrow,\downarrow$ corresponds to up and down spin quantized along the crystalline z-axis.  The positions of the sites within each unit cell $\vec r^l$ can be generated by the SG elements, starting from $\vec r^1 = (1/8,1,1/4)$.
The tight-binding Hamiltonian is given by
\begin{equation}\begin{split}\label{eq:199TB}
H =&   f_{\vec x}^{4 s' \dagger} 
\left( t \, \delta _{s' s} + i \lambda \,  (\sigma^z)_{s' s} \right) f_{\vec x}^{1 s } + \text{h.c.}\\
&  + \text{(symmetry-related terms)}  ,
\end{split}\end{equation}
where `h.c.' denotes Hermitian conjugate, and repeated indices are summed over. In particular, $H$ is TR-symmetric when the parameters $\lambda, t$ are real, and the `symmetry-related terms' include all terms generated under SG symmetries. 
Note that in writing Eq.~\eqref{eq:199TB} we have made a particular choice of the unit cell, and a more explicit form of the Hamiltonian and the symmetry transformation of the sites are provided in Appendix~\ref{sec:a199}.

As shown in Fig.~\ref{fig:199}c, when $\lambda/t=1/4$, the system is a band insulator at filling $\nu = 4$.
This simple observation alone establishes existence of feQBIs.

Before proceeding we note the following points. Without SOC ($\lambda=0$), the system at $\nu=4$ is semi-metallic (Fig.~\ref{fig:199}d). This is in fact the manifestation of a more general result: no feQBI is allowed when spin-rotation invariance is restored~\cite{us3}. 
Interestingly, the lower four bands are also completely flat, since Eq.~\eqref{eq:199TB} describes a nearest-neighbor tight-binding model defined on the frustrated hyper-kagome lattice. In contrast to the flat bands of, say, the kagome lattice, here they cannot correspond to localized, SG-symmetric orbitals due to the non-atomic filling.

\begin{figure}[h!]
\includegraphics[width=\linewidth]{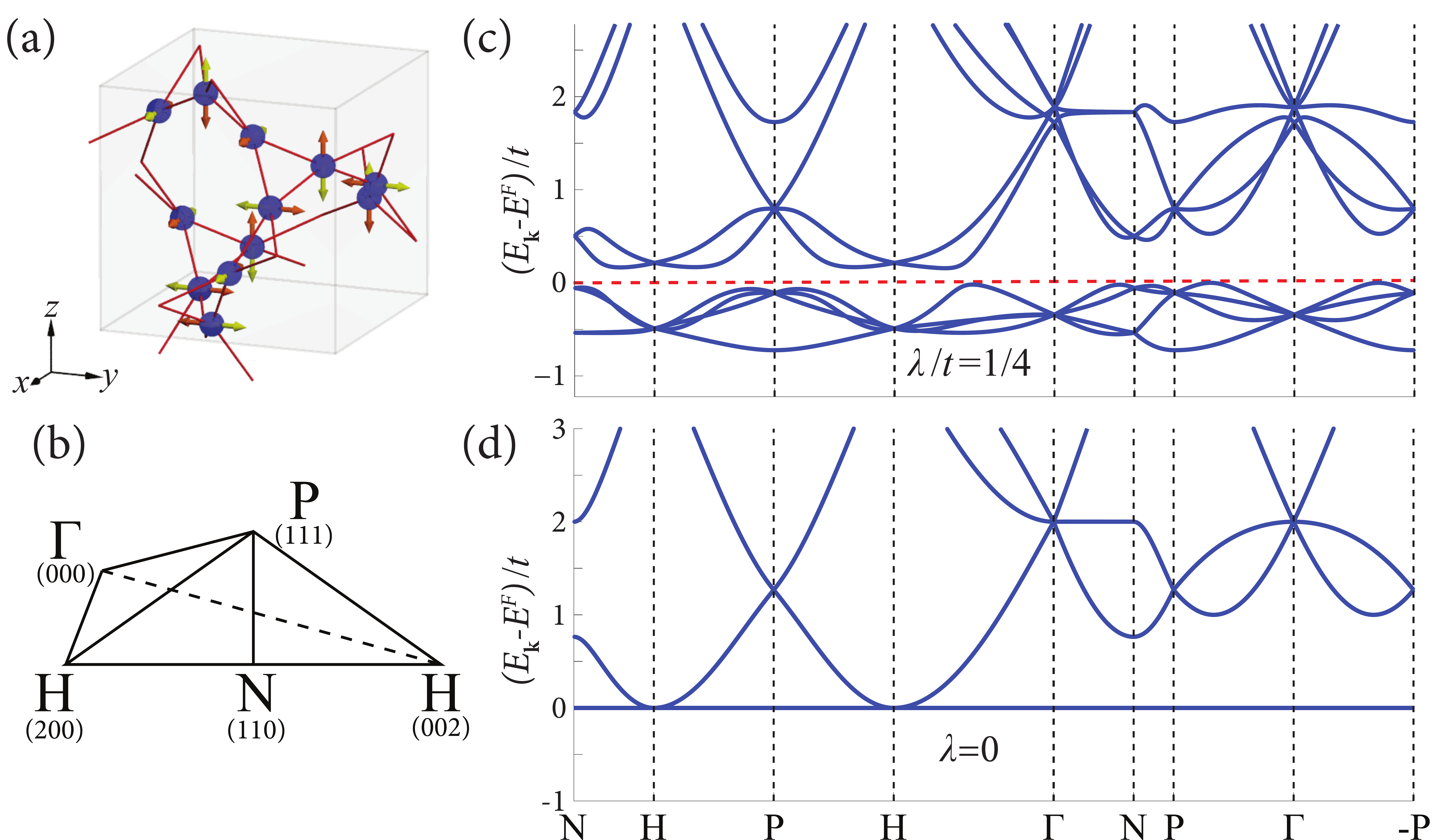}
\caption{
feQBI example with space group No.~199. (a) Hyper-kagome lattice with nearest-neighbour bonds considered in Eq.~\eqref{eq:199TB}.
Spatial symmetry-respecting spin quantization axes can be assigned on each of the sites, which we indicate by colored arrows. 
(b) Irreducible Brillouin zone of the space group, for which any momentum in the first Brillouin zone is symmetry-related to a point inside. Momenta are given in units of $\pi$, setting the lattice constants to $1$. 
(c,d) Example band structures for the tight-binding model in Eq.~\eqref{eq:199TB}.  $E^F$ denotes the Fermi level at filling $\nu=4$.
With spin-orbit coupling in (c), the system is insulating and forms a feQBI. When spin-orbit-coupling is switched off in (d), the lowest four bands are completely flat, but they touch the upper bands at H and the system becomes semi-metallic.  
\label{fig:199}}
\end{figure}

\section{Entanglement spectrum of the `spin-orbital entanglement' cut}
\label{sec:Spin_Ent}
We have established that an insulator at filling $\nu = 4$ in No. 199  cannot be deformed to an AI without breaking any symmetries (even when gap-closing is allowed), and by contrasting with the case of spin-rotation invariant systems one sees that SOC is essential for feQBI.
How does SOC enable such band insulators at non-atomic fillings? Here, we argue that the feQBI has irremovable entanglement between two spin `species', grouped in a way that each spin species is closed under spatial symmetries but interchange under TR. Such entanglement is irremovable as long as the filling is fixed and symmetries are respected, and hence serves as a complementary view on the symmetry obstruction to defining any AI at the same filling.

A standard technique to probe entanglement between degrees of freedom (dof) is via an entanglement cut and its corresponding entanglement spectrum. 
An entanglement cut is a partition of the system into two subsystems $A$ and $\bar A$, corresponding to Hilbert spaces $\mathcal H_A$ and $\mathcal H_{\bar A}$ such that the Hilbert space of the full system `factorizes': $\mathcal H = \mathcal H_{A} \otimes \mathcal H_{\bar A}$.
Most familiarly, $A$ and $\bar{A}$ are spatial regions, though other partitions are possible~\cite{Qi, Israel, Yang, Lundgren2014,BernevigFQHE}.
If $A$ is entangled with $\bar{A}$, then the reduced density matrix $\rho_A = \mbox{Tr}_{\bar{A}} (\ket{\Psi} \bra{\Psi}) $ will be mixed. 
The spectrum of $\rho_A$, often expressed via the `entanglement Hamiltonian' $\rho_A \equiv e^{-H_E}$, contains universal fingerprints of quantum order.
A TI has an intriguing gapless entanglement spectrum at a spatial cut. \cite{Ari}
In contrast, because an AI can  be adiabatically deformed into an un-entangled state by strongly localizing the orbitals, its `entanglement gap' can be deformed to infinity while keeping all symmetries.

The SG symmetries protecting feQBI, however, are broken by any naive spatial cuts. To make maximal use of the symmetries, we introduce a `spin-orbital entanglement cut' (SE cut) that both respect all symmetries and directly probe the irremovable entanglement between spin-species.
The full Hilbert space is a product over the Fock spaces $\mathcal H^F_{\vec x l \sigma}$ of $f_{\vec x l \sigma}$, where $\vec x$ labels spatial location, $l$ labels orbitals and $\sigma = {\Uparrow}, {\Downarrow}$ labels the spin states, i.e. $\mathcal H = \bigotimes_{\vec x l \sigma} \mathcal H^F_{ \vec x l \sigma}$. 
Here we let subsystem $A$ be the collection of spin `up' electrons instead of a spatial region, and the definition of $A$ can be further engineered to respect all the symmetries at hand. 
The entanglement spectrum of the SE cut can be found by tracing over, say, the occupancy states $|n\rangle_{\vec x l {\Downarrow}}$.

We would like all the symmetries `$g$', including the SG, to generate symmetries $U_g$ for the entanglement Hamiltonian $H_{\text{SE}}$, i.e. $U^\dagger_g H_{\text{SE}} U_g = H_{\text{SE}}$ (Appendix \ref{sec:aSpinEnt}). Since electron spin also transforms under spatial symmetries, one must choose the spin quantization axes in a symmetric fashion. Whether such choices exist or not depend on both the SG and the lattice realization, and we present a detailed discussion on this in Appendix \ref{sec:aSpinEnt}. 
Here, we simply note that such SG-symmetric SE cut is possible for SG No.~199 (Fig.~\ref{fig:199}a).
In the following we let $A = {\Uparrow}$ denote the set of dof in the spin-up sector defined by the SE cut, and $\bar{A}={\Downarrow}$ its complement (analogous to `left' and `right' in a typical spatial entanglement cut).

TR exchanges $\Uparrow, \Downarrow$, so its role here is analogous to spatial inversion for spatial entanglement cut~\cite{Ari}.
As detailed in Appendix \ref{sec:aSpinEnt}, TR is realized as a \emph{unitary} symmetry on $H_{\text{SE}}$: while it is anti-unitary, the exchange of the two sub-systems introduces an additional complex conjugation in its proper definition.
The particle number $N$ is also a good quantum number of $H_{\text{SE}}$, and the sum of charges of the two TR-paired Schmidt states is fixed by the total electron number.
Consequently TR is manifested as a particle-hole like symmetry.

With the symmetries of $H_{\text{SE}}$ in hand, we now specialize our discussion to free electrons.
The entanglement Hamiltonian $H_{\text{SE}}$ of a free  system is also free~\cite{Ari, Qi, Peschel, BernevigPRB}, and can be reduced to the single-particle entanglement Hamiltonian $h_{\text{SE}} = \log(C^{-1} - 1)$, where $C_{ij} = \langle \Psi | f^\dagger_{i} f_{j}| \Psi\rangle$ is the correlation matrix and and $i,j$ are restricted to dof in ${\Uparrow}$, and hence run over all of space. Note that $C$ is nothing but the projector onto occupied bands further projected onto the $\Uparrow$ dof.
In contrast to a spatial cut, which introduces a boundary, $h_{\text{SE}}$ behaves like a local Hamiltonian throughout the entire bulk.

By construction $h_{\text{SE}}$ respects \emph{all} the SG symmetries.
In particular, it is periodic and hence one can discuss the entanglement energy bands $\varepsilon^{(l)}_{\text{SE}}(\vec k)$, where $l$ labels the bands.
In addition, $\varepsilon^{(l)}_{\text{SE}}(\vec k)$ are continuous since the physical ground state corresponds to a band insulator.
At physical filling $\nu$, it is straightforward to verify that only the lowest $\nu$ of the entanglement bands can have $\varepsilon^{(l)}_{\text{SE}}(\vec k)<+\infty$, so in what follows we consider only the lowest bands $1 \leq l \leq \nu$.
$H_{\text{SE}}$ has a gapped unique ground state if and only if $h_{\text{SE}}$ is gapped about the entanglement Fermi level $\varepsilon^{\text{F}}_{\text{SE}} = 0$.

For a strict AI, TR is realized locally so $h_{\text{SE}}(\vec k)$ consists of exactly $\nu/2$ bands with $\varepsilon_{\text{SE}}(\vec k)$ at  $-\infty$, and all other bands are at $+ \infty$. 
In particular, the many-body entanglement ground state, constructed by filling all the $\nu_{\text{SE}}= \nu/2$ bands below $\varepsilon^{\text{F}}_{\text{SE}}$, is gapped and unique. The infinite entanglement gap is consistent with the intuition that there is no spin-orbit entanglement in this limit.
Away from the strict atomic limit,  $h_{\text{SE}}$ always \emph{could} become gapless at $\varepsilon^{\text{F}}_{\text{SE}}$, even for an AI, but the question is whether $h_{\text{SE}}$ \emph{must} be gapless.
In the following we argue that the $\nu = 4$ feQBI symmetric under SG No.~199 necessarily has a gapless $h_{\text{SE}}$.

We argue by contradiction.
Assume $H_{\text{SE}}$ is gapped with a unique ground state.
Because TR acts as a particle-hole symmetry, the filling of the entanglement ground state must be half the physical filling, $\nu_{\text{SE}} = \nu / 2  = 2$, implying  bands 1, 2 lie isolated below the Fermi level: $\varepsilon^{(1)}_{\text{SE}}(\vec k) \leq \varepsilon^{(2)}_{\text{SE}}(\vec k) < 0 < \varepsilon^{(3)}_{\text{SE}}(\vec k) \leq \varepsilon^{(4)}_{\text{SE}}(\vec k)$.
However, we find that at $\vec k = \text{P} \equiv (\pi, \pi, \pi)$  the four bands must transform under a 1D and 3D irreps.
It follows from the fact that the collection of irreps at $\text{P}$ carried by the $\nu$ entanglement bands must be the \emph{same} as those of the $\nu$ physical bands.
This is because $h_{\text{SE}}$ is unitarily related to $C$,  the occupied band projector further projected onto the $\Uparrow$ space; the latter projection respects the SG symmetries and so leaves irreps unchanged.
The feQBI here has one 1D irrep and one 3D irrep, and this is a robust property of the system: the counting 4 = 1+3 cannot be altered without closing  the gap, and is independent of any particular tight-binding model used.
This forces $\varepsilon^{(2)}(\text{P})=\varepsilon^{(3)}(\text{P})$, which is incompatible with an entanglement gap and hence a contradiction.

\begin{figure}[h!]
\includegraphics[width=\linewidth]{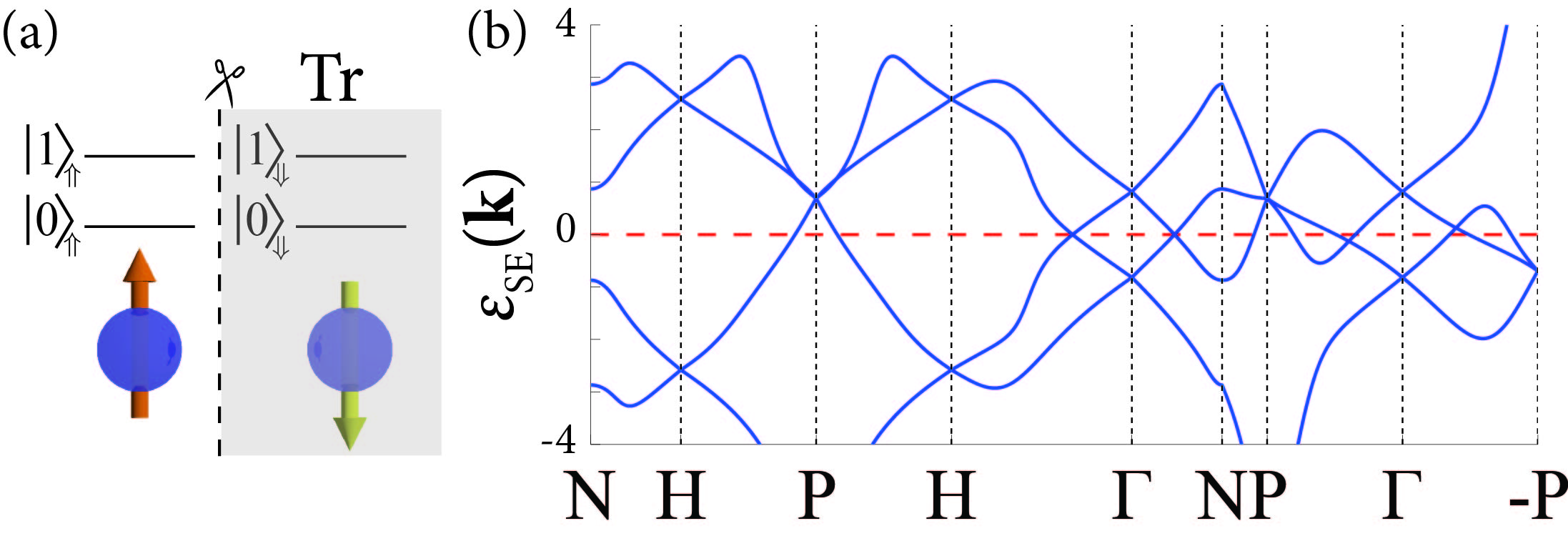}
\caption{
Spin-orbital entanglement cut and entanglement spectrum. (a) Schematic of the entanglement cut, in which the occupancy of one spin species (say ${\Downarrow}$)  is traced over.
(b) Example spin-orbital entanglement cut spectrum for the feQBI model of space group No.~199. The entanglement band structure inherits little group representations from the physical bands, and time-reversal symmetry is realized in a `particle-hole' manner. 
Note that at TR invariant momenta, different irreps of the physical bands can be forced to be degenerate due to TR pairing. Such degeneracies can be lifted in the entanglement bands, as seen at $\Gamma$ and $N$ of the example we gave.
These together force an unavoidable gaplessness about $\varepsilon^F$ (dashed line).
\label{fig:199Ent}}
\end{figure}

\section{Implications and Future Directions}
\label{sec:implications}
While we have focused on SG No.~199 to illustrate the existence and novelty of feQBIs, it is natural to ask if feQBI can exist in systems with other SG symmetries. 
In a recent work~\cite{us} we derived a general lower bound on the minimal filling for all SGs, but the bounds reported there are not necessarily tight. For free electron problems,
tight bounds can be found by symmetry analysis of possible band structures.
We have carried out a comprehensive investigation on the possibility of hosting feQBIs for all SGs, and the results will be presented elsewhere~\cite{us3}.
Here we simply note that, in fact, with SOC $\nu^{\text{Band}}_{\text{min}} = \nu^{\text{AI}}_{\text{min}}$  for \emph{all but the four} Wyckoff-mismatched SGs: No.~199, 214, 220 and 230. These four SGs admit feQBI at $\nu^{\text{Band}}_{\text{min}} = 4,4,8$ and $8$ respectively.
Interestingly, it appears that feQBI arises from the interplay between Wyckoff-mismatch and the spin-1/2 nature of electrons. Understanding this link is an interesting problem we leave for future investigation.

\begin{table}[h!]
\caption{\label{tab:WyckoffPos}
feQBIs in the four Wyckoff-mismatched space groups. `Wyckoff multiplicities' denotes the number of lattice points per \emph{primitive} unit cell required to form a space-group symmetric lattice corresponding to one of the Wyckoff positions. $\nu^{\text{AI}}$  and $\nu^{\text{Band}}$ respectively denote the electron fillings consistent with atomic and  band insulators, and their discrepancy corresponds to feQBIs.
}
\begin{center}
\begin{tabular}{c|ccc}
Space group & Wyckoff multiplicities & $\nu^{\text{AI}}$ & $\nu^{\text{Band}}$\\
\hline
No.~199 ($I2_1 3$) & $4, 6, 12$ & $4 \mathbb N \backslash \{ 4 \}$&  $4 \mathbb N$\\
No.~214 ($I 4_1 3 2$) & $4, 4, 6, 6, 8, 12, 12, 12, 24$ &   $4 \mathbb N \backslash \{ 4 \}$ & $4 \mathbb N$ \\
No.~220 ($I \bar 4 3 d$) & $6, 6, 8, 12, 24$ & $4 \mathbb N \backslash \{ 4,8, 20\}$ & $4 \mathbb N \backslash \{ 4\}$ \\
No.~230 ($I a \bar 3 d$)& $8, 8, 12, 12, 16, 24, 24, 48$ & $8 \mathbb N \backslash \{8\}$ & $8 \mathbb N$
\end{tabular}
\end{center}
\end{table}

Conventional wisdom holds that one can study symmetry protected topological phases by finding obstructions to adiabatically deforming the system to an un-entangled product state (like the AI) while respecting all symmetries.
There were known examples of a `filling-enforced' obstruction to such a deformation \emph{if} one insists on using a particular tight-binding realization of the SG~\cite{Itamar}.
These examples, however, are thought to be trivialized once symmetry-allowed modifications to the lattice models are included, equivalent to bringing down dof from infinity.
Our result indicates there are filling-enforced obstructions to such deformation (adiabatic or not) given \emph{only} the symmetries and filling, independent of the lattice or continuum realization.
While the topological class of the feQBI may well fall under established classifications  (say K-theory, which allows for band addition, and field theories, which do not incorporate the filling~\cite{Kitaev, TI2013}), the `filling-enforced' nature of the feQBI appears to be new, and it would be worthwhile to revisit these programs with `filling' in mind.

We close by discussing possible material realization of feQBI. As indicated in this work, one should consider a system with significant SOC in SG No.~199 or 214 with $\nu=4$ (disregarding core electrons tightly bounded to atoms), or in SG No.~220 or 230 with $\nu=8$. In addition, the energy bands relevant near the Fermi energy should contain the correct symmetry irreps needed for realizing feQBI (Appendix~\ref{sec:a199}). Although we do not yet have a concrete realization in mind, here we point out the hyper-kagome iridate Na$_3$Ir$_3$O$_8$ comes close to providing the ingredients needed: in a hypothetical structure promoting the simple cubic bravais lattice to body-centered cubic (by altering atom positions), the relevant TB Hamiltonian can be adiabatically connected to the model given in Eq.~\eqref{eq:199TB} in certain parameter region (Appendix~\ref{sec:aNa338}).

\section{Conclusion}
\label{sec:conclusion}
In conclusion, a SOC electronic system with TR and SG symmetries can form a band insulator at a filling that is incompatible with any AI. 
This also defies the usual perspective that a TI or TCI can be understood as a `twisted' version of an AI, having been rendered non-atomic due to certain patterns of band inversion at high-symmetry momenta~\cite{TI2013}. 
More generally, our result implies that for certain symmetries there are electron fillings for which a symmetry protected topological phase is the \emph{only} option if the system is symmetric, gapped and short-ranged entangled. We also point out materials exhibiting the hyper-kagome structure are potentially experimental playground to realize feQBIs.

The spin and orbital dof are intrinsically entangled in a feQBI. This was exposed through a `spin-orbital entanglement cut', which can be engineered to make maximal use of all symmetries and is also applicable to interacting many-body systems. 
Curiously, a given feQBI system can simultaneously be a conventional TI (for instance, the model in Eq.~\eqref{eq:199TB} is a strong TI), but this is not a necessary consequence of the filling-enforced nature of feQBIs (Appendix \ref{sec:aTB}).
It is worthwhile to point out, however, that the Wyckoff-mismatched SGs all have SG symmetries that are broken by the existence of any surface, so finding physical bulk signatures of feQBIs is an important open question.

\begin{acknowledgements}
MZ thanks S. Parameswaran for the introduction to NS space-groups and numerous inspiring conversations. AV thanks Ari Turner and Leon Balents for insightful discussions and support from ARO MURI on topological insulators, grant ARO W911NF-12-1-0461. HCP is supported by a Hellman graduate fellowship and NSF DMR 1206728. We thank D.~Varjas for pointing out some of the feQBI examples we found are in fact also strong or weak topological insulators.
\end{acknowledgements}

\bibliography{references}

\appendix
\section{Symmetries of space group No.~199}
\label{sec:a199}
In this appendix we provide supplementary information on the symmetry properties of SG No.~199, an explicit tabulation of the feQBI tight-binding example discussed in Eq.\,(\ref{eq:199TB}), and also an additional tight-binding example.

\subsection{Symmetries of SG No.~199}
Up to lattice translations, No.~199 contains 12 symmetry elements. We parameterize the SG element $g$ by $ g = \{ R_{g} = I e^{- i \vec \theta_{g} \cdot \vec L} ~|~ \vec t_g \}$, where $I = +1$ ($-1$) for proper (improper) rotation. While these can be easily determined from the information listed in ITC~\cite{ITC}, we tabulate them in Table \ref{tab:Sym199} for completeness (`el'. denotes the symmetry operation indexed by the same number in ITC).

\begin{center}
\begin{table}[h!]
\caption{\label{tab:Sym199}
List of symmetry elements in $\mathcal G / T$  for SG No.~199, parameterized by  $ g = \{ R_{g} = I e^{- i \vec \theta_{g} \cdot \vec L} ~|~ \vec t_g \}$. }	
\begin{tabular}{cccc}
el. & $I$ & $\vec \theta_g$ & $\vec t_g$ \\
\hline
(1) & ~$+$~ & $(0,0,0)$ & $(0,0,0)$ \\
(2) & ~$+$~ & $   \pi (0,0,1)$ & $(\frac{1}{2},0,\frac{1}{2} )$ \\
(3) & ~$+$~ &  $\pi (0,1,0)$ & $(0, \frac{1}{2},\frac{1}{2})$ \\
(4) & ~$+$~ &  $\pi (1,0,0)$ & $(\frac{1}{2},\frac{1}{2},0)$  \\
(5) & ~$+$~ &  $\frac{2\pi}{3} \, \frac{1}{\sqrt{3}}(1,1,1)$ & $(0,0,0 )$\\
(6) & ~$+$~ &  $\frac{2\pi}{3} \, \frac{1}{\sqrt{3}}(-1,1,-1)$ & $(\frac{1}{2},\frac{1}{2},0)$ \\
(7) & ~$+$~ &  $\frac{2\pi}{3} \, \frac{1}{\sqrt{3}}(1,-1,-1)$ & $(\frac{1}{2},0,\frac{1}{2})$ \\
(8) & ~$+$~ &  $\frac{2\pi}{3} \, \frac{1}{\sqrt{3}}(-1,-1,1)$ & $(0,\frac{1}{2},\frac{1}{2})$  \\
(9) & ~$+$~ &  $\frac{4\pi}{3} \, \frac{1}{\sqrt{3}}(1,1,1)$ & $(0,0,0 )$\\
(10) & ~$+$~ & $\frac{4\pi}{3} \, \frac{1}{\sqrt{3}}(1,-1,-1)$ & $(0,\frac{1}{2},\frac{1}{2})$ \\
(11) & ~$+$~ & $\frac{4\pi}{3} \, \frac{1}{\sqrt{3}}(-1,-1,1)$ & $(\frac{1}{2},\frac{1}{2},0)$ \\
(12) & ~$+$~ & $\frac{4\pi}{3} \, \frac{1}{\sqrt{3}}(-1,1,-1)$ & $(\frac{1}{2},0,\frac{1}{2})$  
\end{tabular}
\end{table}
\end{center}

\begin{figure}[h!]
\includegraphics[width=1\linewidth]{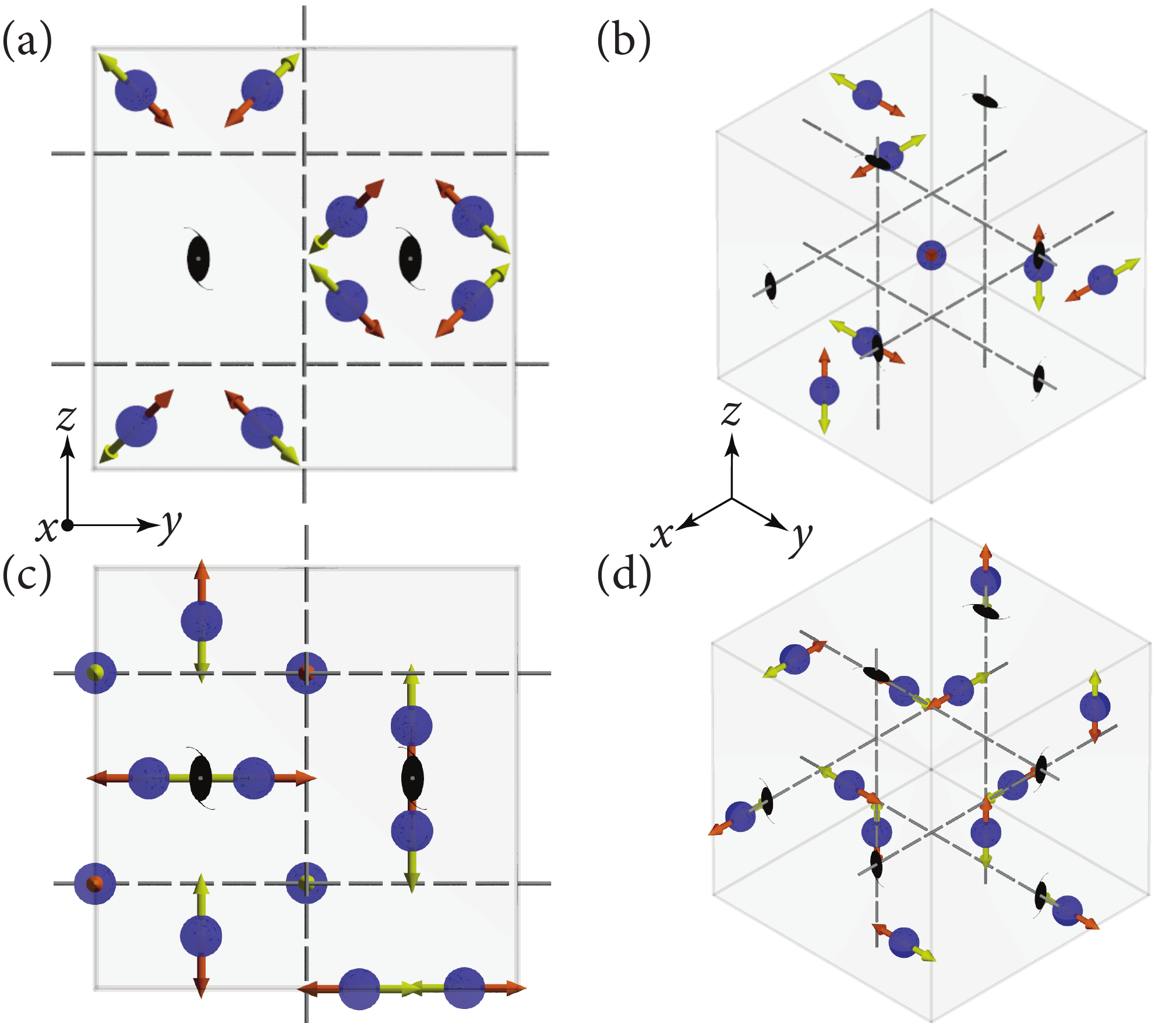}
\caption{ \label{fig:Visualize199}
(a,b) View of representative sites forming $\mathcal W^{\text{199}}_{\text{a}}$ along high symmetry axes. There are 4 sites in each primitive unit cell and hence 8 sites in the conventional cell. (c,d) Same as (a,b) but for $\mathcal W^{\text{199}}_{\text{b}}$. 
}
\end{figure}

SG No.~199 has three Wyckoff positions labeled by a,b and c, which can be realized by having $4$, $6$ and $12$ sites respectively in each primitive unit cell.
To aid visualization of $\mathcal W^{\text{199}}_{\text{a}}$ and $\mathcal W^{\text{199}}_{\text{b}}$, we present in Fig\,\ref{fig:Visualize199} representative lattice realizations of them viewed from different angles.

\subsection{SG symmetric spin-texture}
Sites in $\mathcal W^{\text{199}}_{\text{a}}$ are invariant under a three-fold rotation, but the rotation axis is site-dependent. A SG symmetric spin-texture can be formed by polarizing spins along the corresponding axis on each site. Sites in $\mathcal W^{\text{199}}_{\text{b}}$ are invariant up to lattice translation under one of the two-fold screws, and so the spin-texture has the spin polarized along the corresponding screw axis. The spin-quantization axes are uniquely defined for sites in $\mathcal W^{\text{199}}_{\text{a}}$  and $\mathcal W^{\text{199}}_{\text{b}}$ (up to a global choice of up vs down), but to be self-contained we list them in Table \ref{tab:SpinTexture}.

\begin{table}[h!]
\caption{ \label{tab:SpinTexture}
Spin-quantization axes corresponding to the SG symmetric spin-texture. The sites are labeled by $l$ in the same order as they are listed in ITC~\cite{ITC}. The axes are parameterized as $\hat n_l = (\sin \theta_l \cos\phi_l, \sin \theta_l \sin \phi_l, \cos \theta_l)$, and we denote $\theta_{[111]} = \cos^{-1}(1/\sqrt{3})$.
}
\begin{center}
\begin{tabular}{c|cc}
\multicolumn{3}{c}{$\mathcal W^{\text{199}}_{\text{a}}$ }\\
$l$ & $\theta_l$ & $\phi_{l}$ \\
\hline
1 & $\theta_{[111]}$ & $\pi/4$ \\
2 & $\theta_{[111]}$ & $-3\pi/4$ \\
3 & $\pi-\theta_{[111]}$ & $3\pi/4$ \\
4 & $\pi-\theta_{[111]}$ & $-\pi/4$ \\
\multicolumn{3}{c}{~}\\
\multicolumn{3}{c}{~}
\end{tabular}
~~~
\begin{tabular}{c|cc}
\multicolumn{3}{c}{$\mathcal W^{\text{199}}_{\text{b}}$ }\\
$l$ & $\theta_l$ & $\phi_{l}$ \\
\hline
1 & $\pi/2$ & $0$ \\
2 & $\pi/2$ & $\pi$ \\
3 & $\pi/2$ & $\pi/2$ \\
4 & $\pi/2$ & $-\pi/2$ \\
5 & $0$ & $0$ \\
6 & $\pi$ & $0$
\end{tabular}
\end{center}
\end{table}

We also clarify here a subtle point (though non-essential for our arguments) : if the screws were `intrinsically' NS, no site should be taken back to itself nor its lattice-translation images. This, however, is not the case for SG No.~199 as the screws are not intrinsic: given any of the screws, one can choose an origin such that the `screw' factorizes into a point-group rotation followed by lattice translation. No.~199 is nonetheless non-symmorphic because no common origin can be picked to render all space group elements symmorphic. SGs like No.~199, which are NS but do not contain any intrinsically NS element, are known as `exceptional' NS SGs, and in fact there are only two of them out of the 230 SGs: No.~24 and No.~199.

\subsection{Little group irreducible representations}
In Table \ref{tab:199Rep} we list the symmetry eigenvalues of the different irreducible representations of the little group at high symmetry momenta $\Gamma$, H and P. The set of isolated four bands forming feQBI corresponds to $\rho^{(2b)}_{\Gamma} \oplus \rho^{(2c)}_{\Gamma} $ at $\Gamma$ and $\rho^{(1a)}_{\text{P}}\oplus \rho^{(3a)}_{\text{P}}  $ at P. Irreps at other high symmetry momenta are fixed once these are specified.

\begin{table}[h!]
\caption{ \label{tab:199Rep}
Symmetry eigenvalues of the irreducible little group representations at high-symmetry momenta. We denote $\omega = e^{-i2\pi/3}$.}
(a) $\Gamma = (0,0,0)$  and  $H= (2\pi,0,0)$ :
\begin{center}
\begin{tabular}{c|ccc}
el. & $\rho^{(2a)}_{\Gamma}$ & $\rho^{(2b)}_{\Gamma}$ & $\rho^{(2c)}_{\Gamma}$ \\
\hline
(1) & $\{1, 1\}$ & $\{1, 1\}$ & $\{1, 1\}$\\
(2) & $\{i, -i\}$ & $\{i, -i\}$ & $\{i, -i\}$\\
(3) & $\{i, -i\}$ & $\{i, -i\}$ & $\{i, -i\}$\\
(4) & $\{i, -i\}$ & $\{i, -i\}$ & $\{i, -i\}$\\
(5) & $\{-\omega, -\omega^*\}$ & $\{-1, -\omega\}$ & $\{-1, -\omega^*\}$\\
(6) & $\{-\omega, -\omega^*\}$ & $\{-1, -\omega\}$ & $\{-1, -\omega^*\}$\\
(7) & $\{-\omega, -\omega^*\}$ & $\{-1, -\omega\}$ & $\{-1, -\omega^*\}$\\
(8) & $\{-\omega, -\omega^*\}$ & $\{-1, -\omega\}$ & $\{-1, -\omega^*\}$\\
(9) & $\{\omega, \omega^*\}$ & $\{1, \omega^*\}$ & $\{1, \omega\}$\\
(10) & $\{\omega, \omega^*\}$ & $\{1, \omega^*\}$ & $\{1, \omega\}$\\
(11) & $\{\omega, \omega^*\}$ & $\{1, \omega^*\}$ & $\{1, \omega\}$\\
(12) & $\{\omega, \omega^*\}$ & $\{1, \omega^*\}$ & $\{1, \omega\}$
\end{tabular}
\end{center}

(b) $\text{P}= (\pi,\pi,\pi)$:
\begin{center}
\begin{tabular}{c|cccc}
el. & $\rho^{(1a)}_{\text{P}}$ & $\rho^{(1b)}_{\text{P}}$ & $\rho^{(1c)}_{\text{P}}$ & $\rho^{(3a)}_{\text{P}}$ \\
\hline
(1) & $1$ & $1$ & $1$ & $\{1, 1, 1\}$\\
(2) & $-1$ & $-1$ & $-1$ & $\{-1, 1, 1\}$\\
(3) & $-1$ & $-1$ & $-1$ & $\{-1, 1, 1\}$\\
(4) & $-1$ & $-1$ & $-1$ & $\{-1, 1, 1\}$\\
(5) & $-1$ & $-\omega$ & $-\omega^*$ & $\{-1, -\omega, -\omega^*\}$\\
(6) & $-1$ & $-\omega$ & $-\omega^*$ & $\{-1, -\omega, -\omega^*\}$\\
(7) & $-1$ & $-\omega$ & $-\omega^*$ & $\{-1, -\omega, -\omega^*\}$\\
(8) & $-1$ & $-\omega$ & $-\omega^*$ & $\{-1, -\omega, -\omega^*\}$\\
(9) & $1$ & $\omega^*$ & $\omega$ & $\{1, \omega, \omega^*\}$\\
(10) & $-1$ & $-\omega^*$ & $-\omega$ & $\{-1, -\omega, -\omega^*\}$\\
(11) & $-1$ & $-\omega^*$ & $-\omega$ & $\{-1, -\omega, -\omega^*\}$\\
(12) & $-1$ & $-\omega^*$ & $-\omega$ & $\{-1, -\omega, -\omega^*\}$
\end{tabular}
\end{center}
\end{table}

\section{feQBI tight-binding examples}
\label{sec:aTB}
\subsection{Model given in Eq.~\eqref{eq:199TB}}
One can deduce the transformation of tight-binding (TB) dof by assigning them to some actual locations in real space. In Eq.\,(\ref{eq:199TB}), we assume the sites have coordinates in $\mathcal W^{\text{199}}_{\text{b}}$ at $x=1/8$, with a single $s$-orbital on each site. Note that the TB model is chosen such that it contains the irreps needed to realize feQBI, but feQBI is possible in any TB models that contains the irreps required.

We label the sites (1 - 6) in the same order as they appear in ITC~\cite{ITC} (left - right). In writing their coordinates as $\vec r = \vec x + \vec r^{l}$ with $\vec x $ a lattice vector, $\vec r^l$ depends on the choice of the unit cell. Here, we take the primitive lattice vectors as $\vec a_1 = (1/2, 1/2,1/2)$, $\vec a_2 = (0,1,0)$ and $\vec a_3 = (0,0,1)$, and choose $\vec r^{l}$ such that $\vec r^{l}$ are contained inside the parallelepiped defined by the three primitive lattice vectors.
With this choice of unit cell, the coordinates are given by
\begin{equation}\begin{split}\label{eq:Coor}
\begin{array}{ll}
\vec r^{1} = (1/8,1,1/4); & \vec r^{2} = (3/8,1,3/4);\\ 
\vec r^{3} = (1/4,9/8,1); & \vec r^{4} = (1/4,7/8,1/2);\\
\vec r^{5} = (0,1/4,1/8); & \vec r^{6} = (0,3/4,3/8).
\end{array}
\end{split}\end{equation}
 The transformation of the TB basis is parameterized by writing $g(\vec r^l) = n_i \vec a_i + \vec r^{l'}$, where $\{ n_i\}$ is a triplet of integers. While this is readily computable, in Table \ref{tab:TBTrans} we tabulate the transformation of site $l$ as $(l';~n_1,n_2,n_3)$ under the symmetry elements.
\begin{center}
\begin{table}[h!]
\caption{\label{tab:TBTrans}
Transformation of tight-binding sites, labeled by $l$, under the symmetry elements (el.). $\bar n$ denotes $-n$.
}
\begin{tabular}{c|cccccc}
el.$\backslash$ $l$ &  1 & 2 & 3 & 4 & 5& 6\\
\hline
(1) & $(1;0 0 0)$ & $(2;0 0 0)$ & $(3;0 0 0)$ & $(4;0 0 0)$ & $(5;0 0 0)$ & $(6;0 0 0)$\\
(2) & $(2;0 \bar 2 0)$ & $(1;0 \bar 2 1)$ & $(4;0 \bar 2 1)$ & $(3;0 \bar 2 0)$ & $(5;1 \bar 1 0)$ & $(6;1 \bar 2 0)$\\
(3) & $(2;\bar 1 1 0)$ & $(1;\bar 1 1 0)$ & $(3;\bar 1 1 \bar 1)$ & $(4;\bar 1 1 0)$ & $(6;0 0 0)$ & $(5;0 1 0)$\\
(4) & $(1;1 \bar 2 \bar 1)$ & $(2;1 \bar 2 \bar 2)$ & $(4;1 \bar 2 \bar 2)$ & $(3;1 \bar 2 \bar 2)$ & $(6;1 \bar 1 \bar 1)$ & $(5;1 \bar 1 \bar 1)$\\
(5) & $(3;0 \bar 1 0)$ & $(4;1 \bar 1 0)$ & $(5;2 \bar 1 0)$ & $(6;1 \bar 1 0)$ & $(1;0 \bar 1 0)$ & $(2;0 \bar 1 0)$\\
(6) & $(4;1 \bar 1 \bar 2)$ & $(3;2 \bar 2 \bar 3)$ & $(6;3 \bar 2 \bar 3)$ & $(5;2 \bar 1 \bar 2)$ & $(1;1 \bar 1 \bar 1)$ & $(2;1 \bar 1 \bar 2)$\\
(7) & $(4;0 \bar 1 1)$ & $(3;\bar 1 \bar 1 1)$ & $(5;\bar 1 0 2)$ & $(6;0 \bar 1 1)$ & $(2;0 \bar 1 0)$ & $(1;0 \bar 1 1)$\\
(8) & $(3;\bar 1 0 \bar 1)$ & $(4;\bar 2 1 0)$ & $(6;\bar 2 1 0)$ & $(5;\bar 1 1 0)$ & $(2;\bar 1 0 0)$ & $(1;\bar 1 0 0)$\\
(9) & $(5;2 \bar 1 \bar 1)$ & $(6;2 \bar 1 \bar 1)$ & $(1;2 \bar 1 \bar 1)$ & $(2;1 \bar 1 \bar 1)$ & $(3;0 \bar 1 \bar 1)$ & $(4;1 \bar 1 \bar 1)$\\
(10) & $(6;\bar 2 1 1)$ & $(5;\bar 2 2 1)$ & $(2;\bar 3 2 1)$ & $(1;\bar 2 1 1)$ & $(3;\bar 1 0 0)$ & $(4;\bar 2 1 1)$\\
(11) & $(6;3 \bar 2 \bar 2)$ & $(5;3 \bar 2 \bar 2)$ & $(1;3 \bar 3 \bar 2)$ & $(2;2 \bar 2 \bar 2)$ & $(4;1 \bar 1 \bar 1)$ & $(3;2 \bar 2 \bar 2)$\\
(12) & $(5;\bar 1 0 1)$ & $(6;\bar 1 \bar 1 1)$ & $(2;\bar 2 \bar 1 1)$ & $(1;\bar 1 \bar 1 1)$ & $(4;0 \bar 1 0)$ & $(3;\bar 1 \bar 1 0)$
\end{tabular}
\end{table}
\end{center}

To construct a Hamiltonian $H$ that is fully symmetric under SG $\mathcal G$, one can start from a single (non-symmetric) term $H_0$, like the one given in Eq.~\eqref{eq:199TB}, and take the summation $H=\sum_{g\in \mathcal G} g H_0 g^{-1}$. For reader's convenience, we give an explicit form of $H$ in the momentum space here. To fix notation, we specify a generic term in a periodic Hamiltonian for SOC electrons as follows:
\begin{equation}
\begin{split}\label{eq:TBPar}
H_{\vec{k}} ^{\delta \vec x, l',  t, \vec \lambda,l} =& 
e^{-i\vec{k}\cdot\delta\vec{x}} f_{\vec k}^{l' s'\dagger} (t\,\delta_{s's}+ i (\vec \lambda \cdot \vec \sigma)_{s' s} ) f_{\vec k}^{l s} + \text{h.c.}
\end{split}
\end{equation}
where $\delta \vec x$ is also a lattice vector. In Table \ref{tab:TBSpec}, we specify the Hamiltonian by providing a list of all terms in this notation.
\begin{center}
\begin{table}[h!]
\caption{ \label{tab:TBSpec}
A full list of terms in the feQBI tight-binding example given in the main text. The terms are parameterized as in Eq.\,(\ref{eq:TBPar}).}
\begin{tabular}{ccccc}
 $\vec k \cdot \delta \vec x$  & $~l'~$ & $\Delta$ & $\vec \lambda$ & $~l~$ \\
\hline
 $0$ & $4$ & $t$ & $(0, 0, \lambda)$ & $1$ \\
 $0$ & $3$ & $t$ & $(0, 0, \lambda)$ & $2$ \\
 $0$ & $4$ & $t$ & $(0, 0, -\lambda)$ & $2$ \\
 $-k_z$ & $3$ & $t$ & $(0, 0, -\lambda)$ & $1$ \\
 $\frac{1}{2}(k_x + k_y + k_z)$ & $6$ & $t$ & $(\lambda, 0, 0)$ & $3$ \\
 $\frac{1}{2}(k_x + k_y + k_z)$ & $5$ & $t$ & $(\lambda, 0, 0)$ & $4$\\
 $0$ & $6$ & $t$ & $(-\lambda, 0, 0)$ & $4$ \\
 $k_y + k_z$ & $5$ & $t$ & $(-\lambda, 0, 0)$ & $3$ \\
 $-\frac{1}{2}(k_x + k_y + k_z)$ & $2$ & $t$ & $(0, \lambda, 0)$ & $5$ \\
 $0$ & $1$ & $t$ & $(0, \lambda, 0)$ & $6$ \\
 $-\frac{1}{2}(k_x + k_y + k_z)$ & $2$ & $t$ & $(0, -\lambda, 0)$ & $6$ \\
  $-k_y$ & $1$ & $t$ & $(0, -\lambda, 0)$ & $5$
\end{tabular}
\end{table}
\end{center}

\subsection{An alternative `minimal' model}
Since $\nu^{\text{Band}}_{\text{min}}=4$ for SG No.~199, 	the `minimal' tight-binding model (corresponding to local Hamiltonians) should have at least eight bands. This is indeed possible by considering a pair of orbitals on each of the sites furnishing $\mathcal W^{\text{199}}_{\text{a}}$. However, to match the representation content required for forming a feQBI, the TB dof have to transform in a more intrigue manner under the SG symmetries.

To this end, note again that each site in $\mathcal W^{\text{199}}_{\text{a}}$ is invariant under a three-fold rotation along a site-dependent axis, and hence it can pick-up an orbital phase of $1$, $\omega = e^{-i 2\pi/3}$ or $\omega^2$ under the three-fold rotations. A spin-1/2 polarized along the three-fold rotation axis will also pick up a phase of $e^{ -i 2 \pi /6}$  for spin-$\Uparrow$ and $e^{ i 2 \pi /6}$  for spin-$\Downarrow$. In particular, we consider a TB model formed by the $(\omega,\Uparrow)$ and $(\omega^2, \Downarrow)$ orbitals on each of the symmetry-related sites belonging to $\mathcal W^{\text{199}}_{\text{a}}$. Note that the two on-site orbitals also interchange under TR.

We again parametrize the TB model in a similar manner as in Eq.~\eqref{eq:TBPar}, but here we use an unconventional basis for the Pauli matrices (corresponding to indices $s$ and $s'$), namely we write 
$
\left(
\begin{array}{cc}
f^{l 1 \dagger}_{\vec x} & 
f^{l 2 \dagger}_{\vec x}
\end{array}
\right)
= \left(
\begin{array}{cc}
\tilde f^{l \Uparrow \dagger}_{\vec x}  & 
\tilde f^{l \Downarrow \dagger}_{\vec x}
\end{array}
\right) U(\theta_l,\phi_l)
$, where $\tilde f^{l \Uparrow \dagger}_{\vec x}$ and $\tilde f^{l \Downarrow \dagger}_{\vec x}$ are respectively the creation operators for the $(\omega,\Uparrow)$ and $(\omega^2,\Downarrow)$ orbitals on the $l$-th site in the unit cell with coordinate $\vec x$, and $U(\theta_l,\phi_l)$ is the site-dependent unitary transformation relating the crystalline z-axis to the SG symmetric spin-texture (Table~\ref{tab:SpinTexture}), i.e. had we picked the $\Uparrow$ and $\Downarrow$ states from the same orbital then $f^{ls\dagger}_{\vec x}$ would just correspond to the creation operator written in the basis of crystalline z-axis. 

\begin{figure}[h!]
\includegraphics[width=.8 \linewidth]{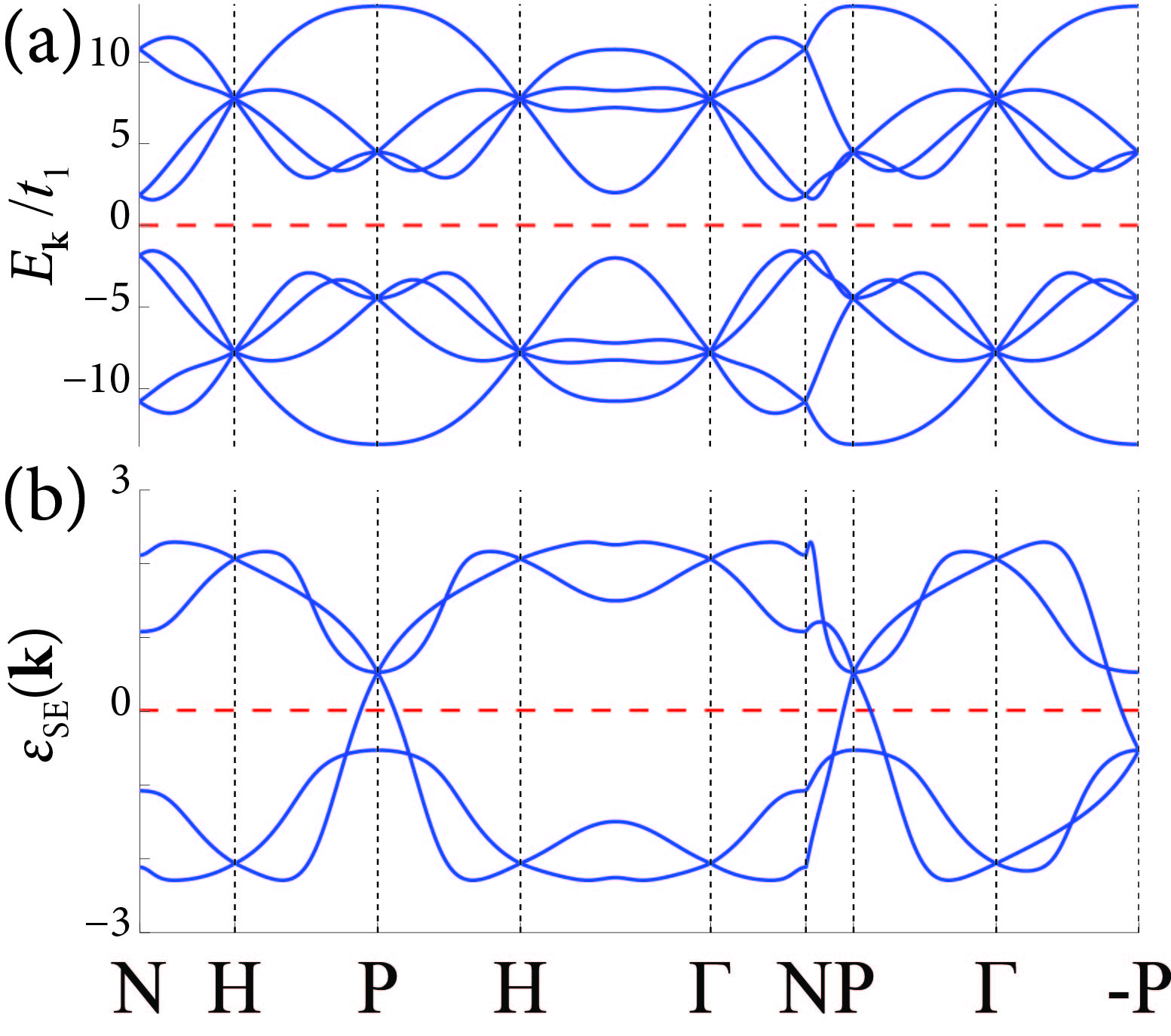}
\caption{
Alternative feQBI example for SG No.~199. Note that the symmetry representation content is identical to the example given in the main text. (a) Band structure with $t_2'/t_1'=2$. Note that the model has an extra `sub-lattice' symmetry (i.e. there exists a unitary $U$ such that  $U H_{\vec k} U = - H_{\vec k}$) and hence the bands are symmetric about $E_{\vec k}=0$, but this is purely an artifact of the simple TB model and is not essential.
(b) Corresponding entanglement band structure.
\label{fig:199_4-4}}
\end{figure}

We consider a TB model with four sites in each unit cell starting with $\vec r^{1} = (0,0,0)$. Similar to the previous example the coordinates $\vec r^{l}$ for $l=1,\dots,4$ can then be determined from the SG symmetries (same choice of unit cell as before), with the sites labeled in the same order as in ITC.
The terms in the eight-band TB Hamiltonian are tabulated in Table \ref{tab:TB2Spec}. In particular, TR invariance of the Hamiltonian still requires $\Delta$, $\vec \lambda$ to be real, although the $\Delta$ term is no longer spin-independent in this choice of basis. One can check that the system forms a feQBI when, say, $t_2'/t_1' = 2$. The corresponding electronic and entanglement bands are plotted in Fig.~\ref{fig:199_4-4}.

\begin{center}
\begin{table}[h!]
\caption{ \label{tab:TB2Spec}
An alternative eight-band feQBI tight-binding example.}
\begin{tabular}{ccccc}
 $\vec k \cdot \delta \vec x$  & $~l'~$ & $t$ & $\vec \lambda$ & $~l~$ \\
\hline
 $0$ & $4$ & $t_1'$ & $(0, 0, 0)$ & $1$ \\
 $0$ & $3$ & $t_1'$ & $(0, 0, 0)$ & $2$ \\
 $0$ & $2$ & $0$ & $(0, t_1', 0)$ & $1$ \\
 $k_y$ & $4$ & $0$ & $(0, t_1', 0)$ & $3$ \\
 $\frac{1}{2}(k_x - k_y - k_z)$ & $3$ & $0$ & $(0, 0, -t_1')$ & $1$ \\
 $\frac{1}{2}(-k_x + k_y - k_z)$ & $4$ & $0$ & $(0, 0, -t_1')$ & $2$ \\
 $-k_z$ & $4$ & $0$ & $(0, 0, t_2')$ & $1$ \\
 $-k_z$ & $3$ & $0$ & $(0, 0, t_2')$ & $2$ \\
 $-k_y$ & $2$ & $0$ & $(-t_2', 0, 0)$ & $1$ \\
 $0$ & $4$ & $0$ & $(t_2', 0, 0)$ & $3$ \\
 $\frac{1}{2}(-k_x - k_y - k_z)$ & $3$ & $t_2'$ & $(0, 0, 0)$ & $1$ \\
 $\frac{1}{2}(k_x + k_y - k_z)$ & $4$ & $t_2'$ & $(0, 0, 0)$ & $2$
\end{tabular}
\end{table}
\end{center}

\subsection{Hyper-kagome lattice and feQBI in the Wyckoff-mismatched space groups No.~214, 220 and 230}
In this subsection we show that the simple TB model in Eq.~\eqref{eq:199TB} can be used to establish the existence of feQBIs in all the four Wyckoff-mismatched SGs: No. 199, 214, 220 and 230. We reproduce in Table \ref{tab:WyckoffPos} the multiplicities of the Wyckoff positions for these four SGs. Note that not all $|\mathcal W^{\mathcal G}_{\text{w}}|$ are integer multiples of $|\mathcal W^{\mathcal G}_{\text{a}}|$ for these SGs.

\begin{figure}[h!]
\includegraphics[width=.8 \linewidth]{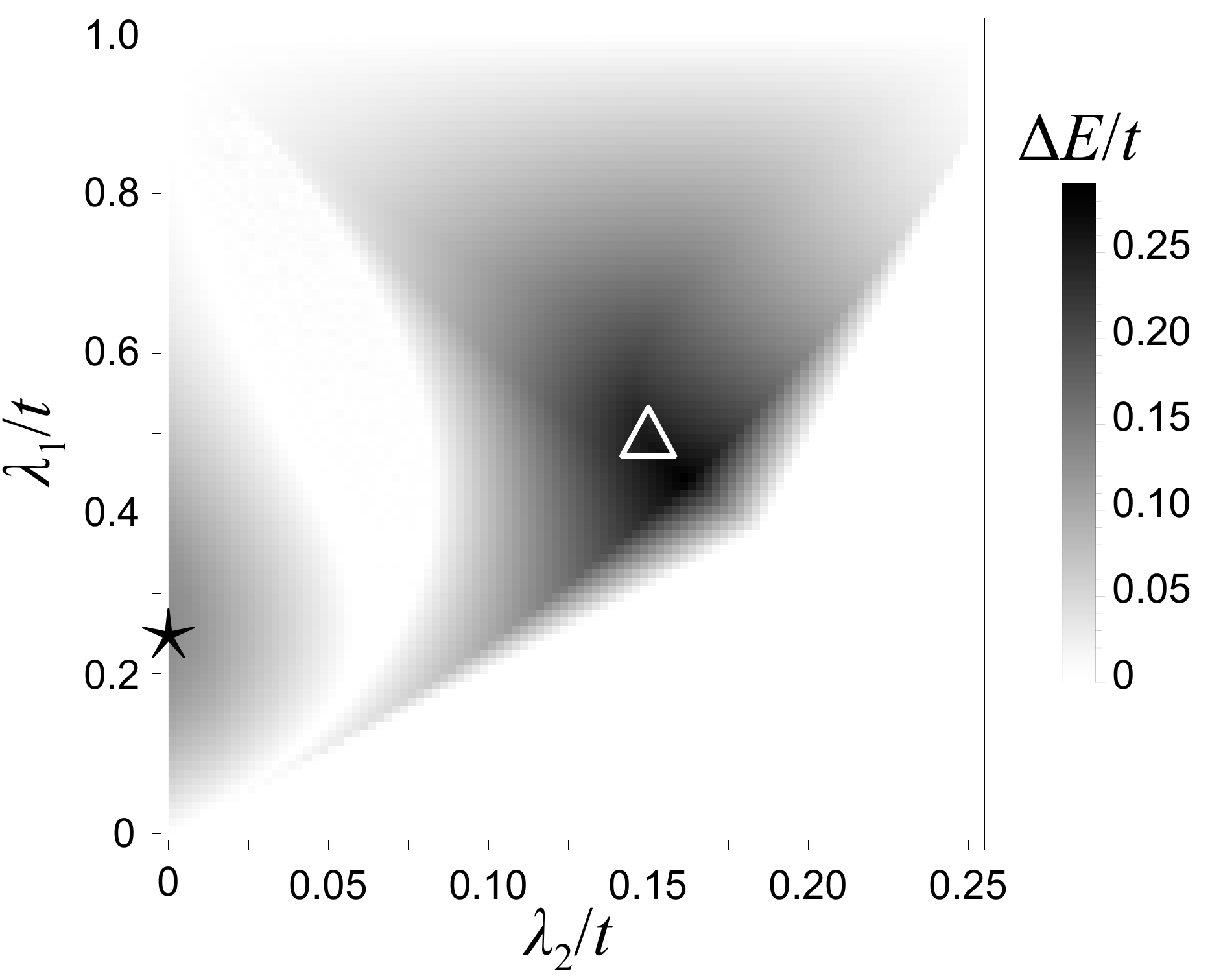}
\caption{
Plot of band gap $\Delta E$ for the hyper-kagome model in Eq.~\eqref{eq:214TB} at filling $\nu=4$. For each pixel of the plot, $\Delta E$ is determined by sampling $30 \times 10^3$ momentum points along the high symmetry lines and in the first Billouin zone respectively. $\star$ corresponds to the system for Fig.~\ref{fig:199}(e,f) of the main text and $\triangle$ corresponds to that of Fig.~\ref{fig:HyperK_Surf}.
\label{fig:HyperK_Phase}}
\end{figure}

To this end, we first focus on SG No.~214. As listed in Table \ref{tab:WyckoffPos}, $|\mathcal W^{\text{214}}_{\text{a}}| = 4$ and so $\nu^{\text{AI}}_{\text{min}} = 8$ for SG No.~214. Now we note that the coordinates listed in Eq.~\eqref{eq:Coor} are chosen to coincide with $\mathcal W_{\text{c}}^{\text{214}}$. Keeping only nearest neighbor bonds, the system is in a hyper-kagome structure (three-dimensional network of corner sharing triangles) and each site is four-coordinated. Demanding TR and spatial symmetries of SG No.~214, the symmetry allowed terms take the form
\begin{equation}\begin{split}\label{eq:214TB}
H =&   f_{\vec x}^{4 s' \dagger} 
\left( t \, \delta _{s' s} + i  (\lambda_1 \, \sigma^z + \lambda_2 \, (\sigma^x-\sigma^y))_{s' s} \right) f_{\vec x}^{1 s }\\
& + \text{h.c.}  + \text{(symmetry-related terms)}.
\end{split}\end{equation}
Eq.~\eqref{eq:199TB} of the main text corresponds to the special case when $\lambda_2=0$, and so it can just as well be regarded as an example of an insulating state in SG No.~214 with filling $\nu=4<8$. This establishes the existence of feQBI for SG No.~214. For completeness, we also plot in Fig.~\ref{fig:HyperK_Phase} the phase of the system described by Eq.~\eqref{eq:214TB} at filling $\nu=4$.

Next we consider the centrosymmetric SG No.~230, which can be viewed as SG No.~214 supplemented with spatial inversion. In particular, $\mathcal W^{\text{230}}_{\text{c}}$ is the union of $\mathcal W^{\text{214}}_{\text{c}}$ and $\mathcal W^{\text{214}}_{\text{d}}$ , which are inversion-copy of each other. Consequentially, one can simply take the TB model in Eq.~\eqref{eq:214TB} together with its inversion copy to construct a TB model symmetric under SG No.~230 (defined on two inter-penetrating hyper-kagome lattices that are decoupled). Hence one can construct a TR symmetric insulator with SG No.~230 at $\nu = 8$. Since $8<2 | \mathcal W^{\text{230}}_{\text{a}}| = 16$, this is also a feQBI.

Finally, we note that SG No.~220 is a subgroup of No.~230. In particular, $8<2 | \mathcal W^{\text{220}}_{\text{a}}| = 12$ and so the same model constructed for No.~230 is also a feQBI example for No.~220.

\subsection{Strong and weak indices of feQBI examples}
The strong and weak indices were found by computing the $\mathbb Z_2$ indices for the six TR-symmetric planes spanned by two of the reciprocal lattice vectors in the Brillouin zone containing either $\Gamma$ or $\vec G_i/2$ (where $\vec G_i$ is the remaining reciprocal lattice vector not spanning the plane). The strong index was found to be $\nu_0 = 1$ and $0$ respectively for the SG No.~199 feQBI models in Eq.~\eqref{eq:199TB} and the one specified in Table \ref{tab:TB2Spec}, while both of them have weak indices characterized by the vector $\vec G_{\nu} = \frac{1}{2}\nu_i \vec G_i =2\pi \, \hat x$. As such, these models actually feature surface states (on the appropriate surfaces), even though they are not required by the filling-enforced nature emphasized in this work.

We note that, however, it is not necessary for a feQBI to simultaneously process nontrivial $\mathbb Z_2$ indices. An explicit example for a feQBI with trivial $\mathbb Z_2$ indices is the inter-penetrating hyper-kagome model constructed for SG No.~220 and 230 (cf.~previous subsection). There the various $\mathbb Z_2$ indices are simply given by twice of the corresponding indices of the model in Eq.~\eqref{eq:199TB}, and therefore must be trivial.

\begin{figure}[htb]
\begin{center}
{\includegraphics[width=1 \linewidth]{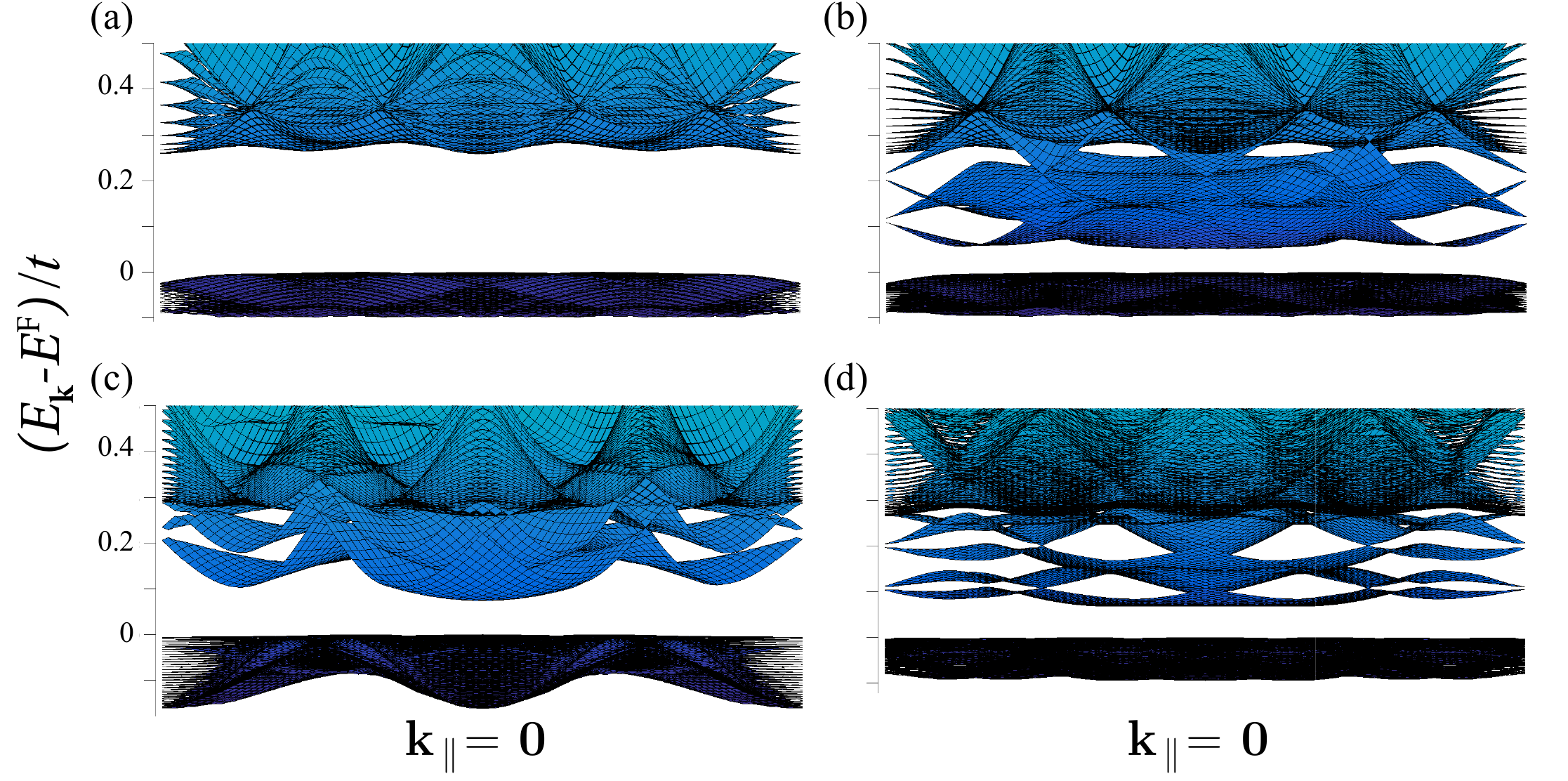}} 
\caption{ Plot of surface band structure against the surface crystal momentum $\vec k_{\parallel}$ for the model in Eq.~\eqref{eq:214TB}. We take $\lambda_1/t = 0.5$ and $\lambda_2/t = 0.15$. $E^{\text{F}}$ denotes the bulk Fermi energy.
Slabs with thickness of $20$ surface-adapted unit cells are used and results for different surfaces, corresponding to different surface normals $\hat n$, are shown. 
Note that with open boundary conditions, there are surface states but they do not traverse the bulk gap.
(a) Periodic boundary condition for $\hat n = \hat z$. (b) Open boundary condition for $\hat n = \hat z$. (c) Open boundary condition for $\hat n \sim \hat y + \hat z$. (d) Open boundary condition for $\hat n \sim \hat x + \hat y + \hat z$.
\label{fig:HyperK_Surf}
}
\end{center}
\end{figure}

As another example, observe that Fig.~\ref{fig:HyperK_Phase} features two `islands' of insulating phases, separated by a gapless region. This is not a coincidence. The island marked by $\star$  corresponds to the strong TI model given in the main text, and the one marked by $\triangle$ is in fact neither a strong nor weak TI. 
As such the mentioned gapless phase between the two islands should actually feature a nodal semi-metal.
We plot in Fig.~\ref{fig:HyperK_Surf} the surface band structures of $\triangle$ on various surfaces, all of which are gapped.

\section{Hypothetical structure for spin-orbit-coupled hyper-kagome material Na$_3$Ir$_3$O$_8$}
\label{sec:aNa338}
In this appendix we point out the observation that the experimentally synthesized hyper-kagome material Na$_3$Ir$_3$O$_8$ (Na-338) is in close proximity to a feQBI phase, and can point to promising avenues for the experimental realization of feQBIs.

Na-338 can be regarded as a hole-doped version of Na-438, a Mott-insulating hyper-kagome material well-known as a spin-liquid candidate~\cite{Balents2010}. In reality Na-338 crystallizes in SG No.~213, which is a primitive version of SG No.~214. The structure of the atoms are listed in Table \ref{tab:Na338_real} (adapted from Ref.~\onlinecite{TakagiNa338}).

\begin{table}[h!]
\caption{Measured structure\cite{TakagiNa338} of Na$_3$Ir$_3$O$_8$ in SG \#213 (Notation as in ITC~\cite{ITC})
\label{tab:Na338_real}
}
\begin{center}
\begin{tabular}{cccc}
Atom & Wyckoff & Free Parameter & Representative\\
\hline
Ir & 12d & $y=-0.113$ & $(0.613, 0.863, 5/8)$\\
O1 & 8c & $x=0.114$ & (0.114, 0.114, 0.114)\\
O2 & 24e & (x,y,z)  & (0.136, 0.907, 0.919)\\
Na1 & 4b & - & $(7/8, 7/8, 7/8)$\\
Na2 & 8c & $x=0.257$ & (0.257, 0.257, 0.257)\\
\end{tabular}
\end{center}
\end{table}

Due to the strong SOC of Iridium and crystal field splitting coming from the local environment, the relevant states near the Fermi level can be (roughly) modeled after an effective $J_{\text{eff}}=1/2$ Kramers pair centered at the Ir sites. 
This gives rise to 24 relevant energy bands around the Fermi level with an electron filling of $\nu=8$.
(Note that in this picture, we are regarding the other 48 bands arising from the $J_{\text{eff}}=3/2$ states of Ir as separated in energy and fully filled, contributing to `core electron states'.)

While Na-338 is semi-metallic in reality~\cite{TakagiNa338}, here we consider a hypothetical structure of it in which the atom positions are modified to `enrich' the spatial symmetries from SG~No. 213 (simple cubic) to No.~214 (body-centered cubic). In particular, we change the free parameter associated with Ir in Table \ref{tab:Na338_real} from $y=-0.113$ to $y=0$, putting the Iridium atoms in $\mathcal W^{\text{214}}_{\text{c}}$. A possible assignment of the other atoms are provided in Table \ref{tab:Na338_hyp}.

\begin{table}[h!]
\caption{`Symmetry-enriched' hypothetical structure of Na$_3$Ir$_3$O$_8$ in SG \#214
\label{tab:Na338_hyp}
}
\begin{center}
\begin{tabular}{cccc}
Atom & Wyckoff & Free Parameter & Representative\\
\hline
Ir & 12c & - & $(0, 1/4, 1/8)$\\
O1 & 8a & - & $(1/8, 1/8, 1/8 )$\\
O2 & 24h &  y= 1/4 & $(0, 1/8, 1/4)$\\
Na & 12d & - & (0, 1/4, 5/8)\\
\end{tabular}
\end{center}
\end{table}

In promoting the bravais lattice from simple cubic in SG No.~213 to body-centered cubic in No.~214, the electron filling per primitive unit cell is halved. Retaining only nearest neighbor bonds, the system can now be effectively described by Eq.~\eqref{eq:214TB} with $\nu=4$ and serves as a feQBI example if the parameters $\lambda_1$ and $\lambda_2$ lie in the insulating phase indicated in Fig.~\ref{fig:HyperK_Phase}.

Note that in this discussion we only intend to give experimental context to feQBI, instead of proposing realistic material candidates. In particular, the `oxygen cage' is significantly distorted from the ideal octahedron form in the structure tabulated in Table \ref{tab:Na338_hyp}, and hence whether the effective spin-1/2 picture still holds or not deserves scrutinization. Nonetheless, we also note that the site-symmetry group for $\mathcal W^{\text{214}}_{\text{c}}$ is the crystallographic point-group $D_2$, which has only one 2-dimensional spinful representation. As such, as long as one can identify a Kramers doublet, well isolated in energy from other states, living on the sites of $\mathcal W^{\text{214}}_{\text{c}}$, the system is described by Eq.~\eqref{eq:214TB} when restricted to nearest neighbor bonds.

\section{Discussions on the spin-orbital entanglement cut}
\label{sec:aSpinEnt}
\subsection{General discussion on time-reversal symmetry}
Consider an electronic system with particle number conservation. Let the total number of particles $N$ be even and $|\Psi\rangle$ be a TR symmetric many-body state. We fix the phase ambiguity such that $\hat {\mathcal T}|\Psi\rangle = | \Psi\rangle$.
Consider a basis $\{ |i\rangle\}$ for sub-system $\Uparrow$ and $\{|\bar i\rangle \} $ for $\Downarrow$ .  In this basis, the ground state can be expanded as
\begin{equation}\begin{split}\label{eq:}
|\Psi \rangle = \sum_{i\bar j} M_{i\bar j} | i\rangle | \bar j \rangle
\end{split}\end{equation}
Generally, $M$ is a rectangular matrix. Since TR exchanges $\Uparrow$ and $\Downarrow$, however, $M$ is in particular square here. Singular-value-decomposition simply gives $M = W \lambda V^{\dagger}$, where $W$ and $V$ are unitary, and $\lambda$ is diagonal and positive semi-definite. The Schmidt states in this basis are given by
\begin{equation}\begin{split}\label{eq:}
|\Psi \rangle = \sum_{\alpha} \lambda_{\alpha} \left( | i\rangle W_{i \alpha}  \right) \left(  | \bar j \rangle V^*_{\bar j \alpha}
\right)= \sum_{\alpha} \lambda_{\alpha} |\alpha\rangle_{\Uparrow} | \alpha\rangle_{\Downarrow},
\end{split}\end{equation}
and in the same basis the reduced density matrices are $\rho_{\Uparrow} = W\lambda^2 W^{\dagger} = M M^{\dagger} $ and $\rho_{\Downarrow} = V^* \lambda^2 V^T = M^T M^*$.

Applying TR operator $\hat {\mathcal T} $, 
\begin{equation}\begin{split}\label{eq:}
| \Psi \rangle = \sum_{\alpha} \lambda_{\alpha} \left( \hat {\mathcal T} |\alpha\rangle_{\Uparrow} \right) \left( \hat {\mathcal T} | \alpha\rangle_{\Downarrow}\right) =\sum_{\alpha} \lambda_{\alpha} \left(  |\alpha_{\hat {\mathcal T}} \rangle_{\Downarrow} \right) \left(  | \alpha_{\hat {\mathcal T} }\rangle_{\Uparrow}\right),
\end{split}\end{equation}
and hence the two Schmidt states $|\alpha\rangle_{\Uparrow}$ and $|\alpha_{\hat {\mathcal T}}\rangle_{\Uparrow}$ are degenerate in the entanglement spectrum. Note that while $\hat {\mathcal T}$ itself is anti-unitary, the map relating the `$\Uparrow$' Schmidt states with the `$\Downarrow$' Schmidt state is also anti-unitary. Altogether, TR becomes a unitary symmetry on the entanglement Hamiltonian.  

To see this more explicitly, let TR transform the basis by
\begin{equation}\begin{split}\label{eq:}
\hat {\mathcal T}| i \rangle = | \bar i \rangle U^{\hat {\mathcal T}}_{\bar i i}; ~~~
\hat {\mathcal T}| \bar i \rangle = | i \rangle \bar  U^{\hat {\mathcal T}}_{i \bar i},
\end{split}\end{equation}
where the unitary matrices $U^{\hat {\mathcal T}}$ and $\bar U^{\hat {\mathcal T}}$ satisfy
\begin{equation}\begin{split}\label{eq:}
\bar U^{\hat {\mathcal T}} U^{\hat {\mathcal T}*}  = (-1)^{\hat N},
\end{split}\end{equation}
as required by $\hat{\mathcal T}^2=(-1)^{\hat N}$. TR invariance of $\hat {\mathcal T} |\Psi\rangle = | \Psi\rangle$ implies
\begin{equation}\begin{split}\label{eq:}
M =&  \bar U^{\hat {\mathcal T}} M^\dagger \left( U^{\hat {\mathcal T} } \right)^{T} ~
\Longrightarrow 
\rho_{\Uparrow}= U^{\hat {\mathcal T}} \rho_{\Downarrow}^* U^{\hat {\mathcal T}\dagger}.
\end{split}\end{equation}
One can bring $\rho_{\Downarrow}$ back to $\rho_{\Uparrow}$ using a `spectral flattened' version of $M$: $\tilde M = W V^\dagger$. See that
\begin{equation}\begin{split}\label{eq:}
\tilde M^{\dagger} \rho_{\Uparrow} \tilde M = W^* \lambda^2 W^T = \rho_{\Downarrow}^*,
\end{split}\end{equation}
so altogether
\begin{equation}\begin{split}\label{eq:}
\rho_{\Uparrow}= (U^{\hat {\mathcal T}} \tilde M^{\dagger}) \rho_{\Uparrow} (\tilde M   U^{\hat {\mathcal T}\dagger}),
\end{split}\end{equation}
i.e. TR is realized as a unitary symmetry on the entanglement Hamiltonian, as claimed.

In particular, since $[ \hat N, \hat {\mathcal T}]=0$, we also have
\begin{equation}\begin{split}\label{eq:}
\hat N \left(  | \alpha_{\hat {\mathcal T} }\rangle_{\Uparrow}\right) = \hat{ \mathcal T} \hat N | \alpha \rangle_{\Downarrow} = (N-N_{\alpha \Uparrow}) \left( | \alpha_{\hat {\mathcal T} }\rangle_{\Uparrow}\right) ,
\end{split}\end{equation}
and hence the charges of the paired states are symmetric about $N/2$, i.e.\,TR is manifested in a `particle-hole' manner. Note that since TR is now realized unitarily, a state with charge $N/2$ can be paired with itself under TR.

To illustrate the ideas discussed above, we provide here a simple illustrative example.
Consider a two site system with a particular TR symmetric state
\begin{equation}\begin{split}\label{eq:}
|\Psi \rangle =& \left ( \alpha  \fd{1}{\uparrow}  \fd{1}{\downarrow}  + \beta ( \gamma \fd{1}{\uparrow} \fd{2}{\downarrow}  - \gamma ^* \fd{1}{\downarrow} \fd{2}{\uparrow}  ) \right.\\
&~~~~\left. +\delta (\epsilon \fd{1}{\uparrow}\fd{2}{\uparrow} + \epsilon^* \fd{1}{\downarrow}\fd{2}{\downarrow} )
\right) | 0 \rangle,
\end{split}\end{equation}
and under TR, $\fd{l}{\uparrow} \rightarrow \fd{l}{\downarrow}$ and $\fd{l}{\downarrow} \rightarrow -\fd{l}{\uparrow}$. Hence,
\begin{equation}\begin{split}\label{eq:}
\hat{\mathcal T }|\Psi \rangle =& \left ( \alpha^*  \fd{1}{\uparrow}  \fd{1}{\downarrow}  + \beta^* (   \gamma \fd{1}{\uparrow} \fd{2}{\downarrow} -\gamma^* \fd{1}{\downarrow} \fd{2}{\uparrow} )\right) | 0 \rangle\\
&~~~~\left. +\delta^* (\epsilon \fd{1}{\uparrow}\fd{2}{\uparrow} + \epsilon^* \fd{1}{\downarrow}\fd{2}{\downarrow} )
\right) | 0 \rangle.
\end{split}\end{equation}
Choosing the global phase such that $\hat{\mathcal T }|\Psi \rangle  = | \Psi\rangle$, we take $\alpha, \beta, \delta \in \mathbb R$.

See that 
\begin{equation}\begin{split}\label{eq:}
_{\downarrow}\langle 0 | \Psi\rangle=& \delta \epsilon\fd{1}{\uparrow}\fd{2}{\uparrow}  | 0 \rangle_{\uparrow} ;\\
_{\downarrow}\langle 0 | \f{1}{\downarrow} |\Psi \rangle =& \left ( -\alpha  \fd{1}{\uparrow} - \beta  \gamma^* \fd{2}{\uparrow} \right) | 0 \rangle_{\uparrow};\\
_{\downarrow}\langle 0 | \f{2}{\downarrow} |\Psi\rangle =& \left (  - \beta  \gamma \fd{1}{\uparrow}\right) | 0 \rangle_{\uparrow};\\
_{\downarrow}\langle 0 |  \f{2}{\downarrow} \f{1}{\downarrow} |\Psi \rangle =& 
 \delta \epsilon^*  | 0 \rangle_{\uparrow},
\end{split}\end{equation}
and so the reduced density matrix after tracing out occupancy states with label $\downarrow$ will have two degenerate eigenvalues of $\delta^2 | \epsilon|^2$. These two Schmidt states, having charge $0$ and $2$ respectively, form a TR pair. The remaining 2-dimensional block is (in the basis of $\fd{1}{\uparrow} | 0 \rangle_{\uparrow}$ and $\fd{2}{\uparrow} | 0 \rangle_{\uparrow}$):
\begin{equation}\begin{split}\label{eq:}
(\rho^{\uparrow})_{2\times 2}=
\left(
\begin{array}{cc}
\alpha^2 + \beta^2 |\gamma|^2 &  \alpha \beta \gamma \\
\alpha \beta \gamma^* & \beta^2 | \gamma|^2
\end{array}
\right).
\end{split}\end{equation}
The corresponding Schmidt weights are
\begin{equation}\begin{split}\label{eq:}
\lambda_{\pm}^2 = \frac{\alpha^2}{2} + \beta^2 |\gamma|^2 \pm \alpha \sqrt{\frac{\alpha^2}{4} + \beta^2 | \gamma|^2},
\end{split}\end{equation}
which are generally non-degenerate, and each state is paired with itself under TR. This is allowed since they have charge $1=2/2$.

\subsection{General discussion on space group symmetries}
The SE cut is defined to partition dof by their spin-label $ {\Uparrow}, {\Downarrow}$, which are picked with respect to some quantization axes.
In the presence of SOC, the spin states on different sites are also related by SG symmetries.
Therefore, to ensure the SE cut is SG symmetric we must carefully choose the spin quantization axes in a site-dependent fashion.

A sufficient condition for the SE cut to respect all SG symmetries is the existence of SG-symmetric spin-texture, i.e. the spins can be polarized in a site-dependent fashion while preserving all the SG symmetries. 
Equivalently, we demand the existence of site-dependent spin-quantization axes for which any spatial symmetries either rotates the spin about the axis, or takes the site to some other site.

Generally speaking, such SG-symmetric spin-textures may not exist for a given lattice realization of a space group: whether or not they can be defined depends on the site-symmetry of the sites in the lattice. 
Since a generic point in space, belonging to the general Wyckoff position, is never taken back to itself under any SG element, the only possible obstruction comes from high-symmetry points (corresponding to high-symmetry Wyckoff positions) where any choice of spin-polarization breaks some site symmetries.

Nonetheless, as long as one is concerned about the space-group symmetries, but not the specific lattice realization of the system, any obstruction to defining a SG symmetric spin texture is largely technical. Intuitively, in the continuum it should be possible to adiabatically `punch-out' the high-symmetry points without changing the phase. More concretely, one can always `split up' a high-symmetry orbital into a small set of orbitals in the vicinity of the high-symmetry point in order to avoid any obstruction.
In other words, one can approximate any desired orbitals on a lattice site by effective `molecular orbitals' on sites belonging to the lower-symmetry Wyckoff positions.

We also note that if the site-symmetry group is one of the 27 non-cubic crystallographic point groups, then all SG symmetries can be regarded as symmetries of the entanglement Hamiltonian (even when SG-symmetric spin-texture cannot be defined).
This is due to the existence of a `primary rotation axis' on the sites: for these 27 point groups, a spin polarized along the primary rotation axis is either flipped or left invariant (up to a phase) by any symmetry operations. A SG symmetry is then realized as an anti-unitary or unitary symmetry on the entanglement Hamiltonain depending on whether it flips the spin or not. In particular, we note that this is true for all the Wyckoff positions for the four Wyckoff-mismatched space groups.

\subsection{Single-particle entanglement Hamiltonian}
In the following subsections we specialize our discussion to free electron problems.
Due to Wick's theorem, the ground state entanglement spectrum of a free fermion system partitioned into $A$ and $\bar A$ is completely captured by the single-particle correlation matrix $C_{ij} = \langle \Psi| f^\dagger_{ i} f_{j} | \Psi \rangle$, where $|\Psi\rangle$ is the many-body ground state and  $i,j, \,\in A$~\cite{Peschel,Ari,Qi,BernevigPRB}. Since the spin-orbital entanglement cut preserves spatial symmetries, and in particular translation invariance, we can focus on correlation functions evaluated at a fixed momentum $\vec k$. This is related to the filled-band projector $\mathcal P_{\vec k}$ by $C_{\vec k; \alpha \beta} = (\mathcal P_{\vec k}^T)_{\alpha \beta}$, where $\alpha, \beta$ are now collective index for basis, orbitals and spin dof, and run from $1, \dots, 2m$	. In a proper choice of basis, the first $m$ entries of $\mathcal P_{\vec k}$ corresponds to the spin `up' sector in the SE cut, and the $m\times m$ restricted correlation matrix is simply 
\begin{equation}\begin{split}\label{eq:CorrH}
C^{\Uparrow}_{\vec k} = 
\left(
\begin{array}{cc}
1_{m\times m} & 0_{m\times m} 
\end{array}
\right)
\mathcal P_{\vec k}^T
\left(
\begin{array}{c}
1_{m\times m} \\
0_{m\times m} 
\end{array}
\right).
\end{split}\end{equation}
$C^{\Downarrow}_{\vec k}$ is similarly defined. The definition is, of course, basis independent - a basis transformation of $\mathcal P_{\vec k}$ is always accompanied by a corresponding transformation of the rectangular matrix projector. Since $h_{\text{SE}}(\vec k)$ is completely determined by $C^{\Uparrow}_{\vec k}$, we focus on the properties of $C^{\Uparrow}_{\vec k}$  in the following.

As $C^{\Uparrow}_{\vec k} $ is a product of projectors, we have $\text{eig}(C^{\Uparrow}_{\vec k} ) \in [0,1]$. A (many-body) Schmidt state $|\lambda\rangle$ with charge $N_\lambda$ is formed by populating $N_\lambda$ single particle eigenstates $| c_i\rangle$ of the full correlation Hamiltonian $C^{\Uparrow} = \bigoplus_{\vec k} C^{\Uparrow}_{\vec k} $. The eigenvalues of the reduced density matrix of such a state is a product over the eigenvalues of the occupied (occ.) and unoccupied (unocc.) states~\cite{Peschel, Ari,Qi,BernevigPRB}:
\begin{equation}\begin{split}\label{eq:}
\lambda^{2} = \prod_{i\in \text{occ.}} c_i \prod_{j\in \text{unocc.}} (1-c_j)
\end{split}\end{equation}
and hence the highest weight Schmidt states are formed by filling all states of $C^{\Uparrow} $ with $c_i >0.5$, i.e. in this language one should think of filling the bands of  $C^{\Uparrow}$ from above, not below, and the effective `chemical potential' is $0.5$. Note that the single-particle entanglement energies are related to $c_i$ here by $\varepsilon_i = \log(c_i^{-1} -1)$.
In particular, if there are states with $c^i = 0.5$, they contribute to $\lambda$ in the same way whether they are filled or not, and corresponds to degeneracy in the entanglement ground state. Note also due to particle-hole character of the TR symmetry on the entanglement spectrum, the highest weight Schmidt state, if unique, must have a charge corresponding to half of the real filling.

\subsection{Symmetry properties of the single-particle entanglement Hamiltonian}
Recall that a single-particle Hamiltonian symmetric under a spatial symmetry $g$ satisfies
\begin{equation}\begin{split}\label{eq:}
U_{\vec k}^{g \dagger }H_{\vec k} U^{g}_{\vec k} =&  H_{g^{-1}( \vec k)} 
\end{split}\end{equation}
where $U^{g}_{\vec k}$ is a unitary. The notion of a spatially symmetric spin-texture implies that on each site, one can specify a spin quantization axis such that the two spin eigenstates are decoupled under all spatial symmetries.
This implies there exists a $\vec k$-independent unitary matrix $\mathcal W_S$ such that 
\begin{equation}\begin{split}\label{eq:SpinTexture_U}
\mathcal W_S  U^{g}_{\vec k} \mathcal W_S ^\dagger =  
\left(
\begin{array}{cc}
U^{g\Uparrow}_{\vec k} & 0\\
0 & U^{g\Downarrow}_{\vec k} 
\end{array}
\right).
\end{split}\end{equation}
For simplicity, in the following we assume we work with such a basis in mind such that we already have $U^{g}_{\vec k} = U^{g\Uparrow}_{\vec k}  \oplus U^{g\Downarrow}_{\vec k} $. Since the band projector satisfies
\begin{equation}\begin{split}\label{eq:}
\mathcal P_{\vec k} =  U^{g \dagger}_{\vec k} \mathcal P_{g^{-1}(\vec k)} U^{g }_{\vec k},
\end{split}\end{equation}
the restricted correlation matrix transforms as
\begin{equation}\begin{split}\label{eq:}
C_{\vec k}^{\Uparrow} 
=
\left( U^{g \Uparrow}_{\vec k} \right)^T
C_{g^{-1}(\vec k)}^{\Uparrow} 
\left( U^{g \Uparrow }_{\vec k} \right)^*
,
\end{split}\end{equation}
i.e. it transforms as a spin-polarized system with the same SG symmetries.

It remains to show how the original TR symmetry is manifested in $C^{\Uparrow}_{\vec k}$. Note that for the original Hamiltonian, TR symmetry implies $H_{\vec k} =  U_{\mathcal T} H_{- \vec k}^*  U_{\mathcal T}^\dagger$, and therefore $\mathcal P_{-\vec k} =  U_{\mathcal T}^{T} \mathcal P_{\vec k}^*  U_{\mathcal T}^{*}$. In the basis of Eq.(\ref{eq:CorrH}), one simply has $ U_{\mathcal T} = -i\sigma^y \otimes 1_{m\times m}$, and therefore $C^{\Uparrow}_{\vec k} = \left( C^{\Downarrow}_{\vec k}\right)^*$. Using the properties of the projectors and the duality discussed in Refs.~\onlinecite{Qi,Yang}, one sees that if $c_{\vec k}^i \in (0,1)$ is an eigenvalue of $H_{\vec k}^{C\Uparrow}$, then $1 - c_{\vec k}^i$ is an eigenvalue of  $C_{-\vec k}^{\Uparrow}$, i.e. TR is now manifested as a `particle-hole' symmetry between $h_{\text{SE}}(\vec k)$ and $h_{\text{SE}}(-\vec k)$ (Fig.~\ref{fig:199Ent}b and~\ref{fig:199_4-4}b). 
More explicitly, suppose $C_{\vec k}^{\Uparrow} \psi = \psi c_{\vec k} $, and let
\begin{equation}\begin{split}\label{eq:}
\tilde \psi=
\left(
\begin{array}{cc}
0 & 1
\end{array}
\right) 
\mathcal P_{\vec k}^T
\left(
\begin{array}{c}
\psi\\
0
\end{array}
\right).
\end{split}\end{equation}
One can then verify $C_{\vec k}^{\Downarrow} \tilde \psi = (1-c_{\vec k}) \tilde \psi$. Hence, as long as $\tilde \psi \neq 0$, it is an eigenvector of $H_{\vec k}^{C\Downarrow} $ with eigenvalue $1 - c_{\vec k}$. In addition, one can check that $\tilde \psi = 0$ implies $c_{\vec k} = 0$ or $1$, so any $c_{\vec k} \in (0,1)$ is paired with a $c_{-\vec k} = 1 - c_{\vec k}$, i.e. translating this relation into the spectrum of $h_{\text{SE}}(\vec k)$, TR is reflected as a particle-hole like symmetry with $\varepsilon_{\text{SE}}(\vec k) = - \varepsilon_{\text{SE}}(-\vec k)$.

\clearpage

\end{document}